# 计算流体力学的时空观: 模型的时空关联性与算法的时空耦合性

IAPCM，北京

# 计算流体力学的时空观：模型的时空关联性及算法的时空耦合性[1]

**李杰权**[2][3]

北京应用物理与计算数学研究所计算物理重点实验室；
北京大学应用物理与技术中心

**摘要：** 流体力学中波的有限传播、粒子的碰撞、各种力之间相互作用，无不体现时空关联效应。本文以计算方法的视角探讨计算流体力学的时空观，即流体力学模型的时空关联性，计算方法的时空耦合性。从流体力学微团法建模出发，明确模型时空关联性的涵义，建立有限体积格式的基本原理，阐述算法时空耦合的必要性，实现流体力学基本控制方程物理建模与有限体积格式数学原理的统一。在实践中，给出时空耦合高精度数值方法设计思路，通过算例比较它与时空解耦方法的差别。需要指出，本文的大部分内容适用于连续介质假定下的计算流体力学研究，部分仅适用于可压缩流动。

## 一 引言

1959 年，华罗庚先生在《大哉数学之为用》（数与量）【1】一文中关于“时空”有以下论述：

*四维空间听来好象有些神秘，其实早已有之，即以“宇宙“二字来说，‘往古来今谓之宙，四方上下谓之宇”（《淮南于 · 齐俗训》)就是宇是东西、南北、上下三维扩展的空间，而宙是一维的时间。*

接着他引用相对论和生活的常识叙述了时空既独立又统一的性质：

*爱因斯坦不再把“宇”、“宙”分开来看，也就是时间也在进行着。每一瞬间三维空间中的物质在占有它一定的位置。他根据麦克斯韦一洛伦兹的光速不变假定，并继承了牛顿的相对性原理而提出了狭义相对论。狭义相对论中的洛伦兹变换把时空联系在一起，当然并不是消灭了时空特点。如向东走三里，再向西走三里，就回到原处，但时间则不然，共用了走六里的时间。时间是一去不复返地流逝着。*

华先生意在说明数学在描述“宇宙之大”的作用，本文的引用旨在强调时空之间的关系深刻地植根于计算流体力学的模型理解和算法构造。离开了时空演化，一切归于“稳态”。正因为时空演化，流体的世界才是色彩斑澜，波澜壮阔。

定性甚至定量描述流体世界的时空演化并不容易，数值模拟同样是貌似简单，实则不易。下面是众所周知的例子，

$$\partial_t u(x,t) + a\,\partial_x u(x,t) = 0, \tag{1}$$

[1] **基金项目：**国家自然科学基金（11771054，91852207），国家重大项目GJXM92579，计算物理重点实验室基金
[2] **作者简介：**李杰权（1969.3），男，安徽，研究员，研究方向：计算流体力学、偏微分方程，数值分析
[3] **通信作者：**李杰权， Email：li_jiequan@iapcm.ac.cn.

其中$a$是一个常数，$x$表示空间坐标，$t$是时间变量。此模型表示了变量$u(x,t)$的空间变化(扰动、波)与时间变化(传播)的关系，描述了一个时空关联的性质

$$u(x,t+\Delta t)=u(x-a\Delta t,t). \tag{2}$$

此表达式事实上比(1)更加根本和直接，并不牵涉函数$u(x,t)$本身更多性质,比如正则性. 即使$u(x,t)$是一个方波或脉冲，(2)仍然有效。用平移算子$e^{-a\Delta t\,\partial_x}$来表示(2)，则有

$$u(x,t+\Delta t)=u(x-a\Delta t,t)=e^{-a\Delta t\,\partial_x}u(x,t). \tag{3}$$

从这里出发，如果想精确表达(1)的输运性质，则需要进行展开(如泰勒展开)

$$\begin{aligned} u(x,t+\Delta t) &= u(x-a\Delta t,t)=e^{-a\Delta t\,\partial_x}u(x,t) \\ &= u(x,t)-a\Delta t\,\partial_x\,u(x,t)+\frac{(-a\Delta t\,\partial_x)^2}{2}u(x,t)+\cdots \end{aligned} \tag{4}$$

为了设计实际工程中的计算方法（简称算法），需进行必要操作：一是截断处理，涉及算法与控制方程的相容性，涉及精度的概念；二是对空间变化$\partial_x u(x,t)$的(如有限差分)处理，不同的处理得到不同的算法。这些都与函数$u(x,t)$的正则性和模型(1)的内在性质息息相关：正则性越好，精度越高；模型(1)的时空关联性需要通过(2)精确描述. 这些处理充满技巧和无限可能，怎样“令人信服”地设计时空耦合算法且精准描述相应模型的内在时空关联性，是一个自然的、但要深入思考的本质问题。

近年来，计算流体力学得到蓬勃的发展，各种算法和相应的应用软件目不暇接。有些算法具有坚实的理论基础，如基于变分原理的有限元方法等。而流体力学中关于可压缩流动的研究，由于流场结构非常复杂(激波、物质界面、涡团和湍流等),人们对流动（解的）性质缺乏很好理解，模型的适定性和相关数值模拟中算法的探讨相当工程化，形式化的表达往往导致似是而非的结果，常常是模型具有很好的性质，而相应数值模拟的离散模型失去了该保持的性质，比如模型的时空关联性和算法的时空耦合性等。不同的出发点导出的数值算法往往截然不同。本文试图在这方面做一点尝试：基于连续介质假定探讨流体力学模型时空变化的内在关联性 (inherent spacetime correlation), 阐述相应算法应保持的时空耦合(spacetime coupling)性质。

本文从流体力学微团法的直接建模开始思考模型的时空关联性。通过对通量的研究，发现经典的瞬时 Cauchy 通量无法反映时空关联性，提出了某一时间段内总通量（lump flux），在任意特定时间段内反映流体的动力学过程【2，3】，进而提出**有限体积格式的基本原理：在连续介质假定性下，任一时间段内流体通过控制体边界的总通量关于边界的扰动是 Lipschitz 连续的**。这一 Lipschitz 连续性恰好体现流场时空关联性的本质特征，实现了数学上流体力学方程组弱解和物理上积分型平衡律(integral balance laws)之间的统一，后者其实就是全离散有限体积格式的物理起源。通过对流体力学方程组时空关联性质的研究，实现有限体积框架下的时空耦合。在算法的实现过程中，**物理量（质量、动量、能量等）以及它的变化率**这一对量(pair)起着重要的作

用，一个量的变化率通过它的空间变化关系来反映。变化率不是一类常规的数学演算，而是反映了内在的物理规律。例如，加速度是速度的变化率，它等于控制体表面受到的力，即应力，这是牛顿第二定律。在流体力学计算方法中使用这样的量是自然的也是必须的，反映了时空变化的联系。在文【4】，笔者倡导的时空耦合高精度算法正是这一思想的具体表现。本文再次描述实现时空耦合算法的一种基本流程，并通过算例展示时空耦合算法的数值表现。

这里倡导的计算流体力学时空观不是新的，从偏微分方程的 Cauchy-Kowalevski 方法到 Lax-Wendroff 方法就是某种意义下的时空耦合方法. 现代计算流体力学的算法时空观常有体现，如对流方程的迎风格式、广义黎曼解法器【5, 6, 7】、守恒元/解元(CE/SE)方法【8】、气体动理学解法器【9, 10】等。最近几年，笔者致力于思考高精度时空耦合算法的必要性与理论基础，但远远不够深入. 随着计算技术的提高和工程需求的精细化，这方面的研究应该得到重视。故作此文，抛砖引玉。

## 二 流体力学方程组的时空关联性

众所周知，流体力学建模过程常用微团法，即在某一时间段$(t, t+\delta t)$，研究控制体$(\boldsymbol{x}, \boldsymbol{x}+\delta \boldsymbol{x})$内流体质团的运动。时间间隔$\delta t$非常重要，给出了微团运动的动力学过程响应时间。为了数学的描述方便，基于连续介质假定，流体力学方程组写成如下形式：给定一个控制体$\Omega(t)$, 流体运动满足下列的关系[4]

$$\frac{d}{dt}\int_{\Omega(\mathrm{t})} \boldsymbol{u}(\boldsymbol{x}, t)\, d\boldsymbol{x} = -\int_{\partial\Omega(t)} \boldsymbol{f}\cdot\boldsymbol{n}\, d\Gamma \tag{5}$$

这里$\boldsymbol{u}$是密度函数，$\boldsymbol{n}$是$\partial\Omega(t)$的单位外法向。相应地，称

$$\boldsymbol{m}(t) = \int_{\Omega(t)} \boldsymbol{u}(\boldsymbol{x}, t)\, d\boldsymbol{x} \tag{6}$$

为控制体$\Omega(t)$上的质量，右边

$$\mathcal{C}(\partial\Omega; t) = -\int_{\partial\Omega(t)} \boldsymbol{f}\cdot\boldsymbol{n}\, d\Gamma \tag{7}$$

为 Cauchy 通量(flux)，$\boldsymbol{f}$为通量密度函数，简称为通量函数，它是$\boldsymbol{u}$以及空间变化量$\nabla\boldsymbol{u}$的函数$f = \boldsymbol{f}(\boldsymbol{u}, \nabla\boldsymbol{u}, \cdots)$。方程组(5)反映了质量时间变化率与通过控制体边界$\partial\Omega(t)$的通量之间的瞬时关系。

Cauchy在特定连续假定下，论述了Cauchy通量的性质【11, 12】，这是他对连续介质力学(Continuum Mechanics)最重要的贡献【11】。要深入理解这一关系并不容易，这与Gauss-Green定理密切相关。基于这一定理，可以将(5)写成散度型偏微分方程(PDE)形式[5]

---

[4] 略去外力场等效应.

[5]略去雷诺输运定理等细致讨论，并设Ω不随流场运动，即欧拉型控制体。相应地，(1)中的控制体为拉格朗日控制体。

$$\partial_t \boldsymbol{u} + \nabla \cdot \boldsymbol{F}(\boldsymbol{u}, \nabla \boldsymbol{u}, \cdots) = 0. \tag{8}$$

这个方程对光滑流动是有效的，可由偏微分方程的知识建立时空的关联性。对于实际流动来说，这一转化是非常粗糙的，难点在于Cauchy通量$\mathcal{C}(\partial\Omega;\mathrm{t})$的数学性质，即正则性。可压缩流场蕴涵丰富的现象，如冲击波、物质界面、湍流等效应，流场（或称方程的解）的正则性非常弱，人们对它的认识很少，故而该定理的应用并不显然。历史上有很多研究【13, 14, 15, 16】，直到最近这方面的研究工作还是热点【17】，遗憾的是所取得的结果与实际的期望仍然相差甚远。正如最近的专著【18, Section 1. 3】有这样一段论述：“*Regarding the right-hand side of (1.3) one needs to keep in mind the following comment concerning the identification of the boundary flux: "the drawback of this functional analytic, demonstration is that it does not provide any clues on how the* $q_D$ *may be computed from A* ”，其中$q_D$对应于这里的 $\boldsymbol{f}(\boldsymbol{u}) \cdot \boldsymbol{n}$， $A$对应于一个给定的应力张量(stress tensor)。也就是说：人们还不知道如何从应力张量中计算出通量函数。

仔细审视可以看出Cauchy通量 $\boldsymbol{\mathcal{C}}(\boldsymbol{\partial\Omega}; \boldsymbol{t})$是表示一个特定瞬时$t$的行为，方程(5)只是表示一个非常弱的“时-空”关联性质。现在常用的一个观点是下面的“弱解”概念，它来自于偏微分方程(8)：

**定义1.1.** 设$\Omega$是$\mathbb{R}^n$中的一个有界区域，$n = 1,2,3$, $0 \le t_1 \le t_2$. 对任意试验函数$\phi(x,t) \in C_0^\infty(\Omega \times [t_1, t_2])$，如果函数$u(x,t)$满足

$$\int_{t_1}^{t_2} \int_\Omega [\boldsymbol{u}(\boldsymbol{x},t)\partial_t \phi + \boldsymbol{F} \cdot \nabla\phi] d\boldsymbol{x} dt = 0, \tag{9}$$

则称$\boldsymbol{u}(\boldsymbol{x},t)$是(8)的弱解。

这个表达式比(5)前进了一大步：如果解$\boldsymbol{u}(\boldsymbol{x},t)$是光滑的，(8)和(9)是等价的。近年来偏微分方程理论得到极大的发展，人们对于(8)的认识比直接对(5)的认识要深刻得多，且偏微分方程(8)的时空关联性一目了然：流场的空间摄动可以通过(8)反映到时间的变化中，线性波的传播(1)正是这种时空关联性的例子。另外，当一个间断面在无限薄的假定下，可以导出Rankine-Hugoniot间断关系，即间断面（线）两侧解的“迹”的关系。

定义1. 1并没有对函数$\boldsymbol{u}(x,t)$做过多假设，只要(9)成立即可。一旦流场出现复杂间断或其它流场结构时，这个定义并不能给出我们太多的信息。我们回避不了的事实是：在可压缩流动中，流场即使初始性质非常“好”，但在不久的将来由于复杂的非线性相互作用，可能变得非常“糟糕”。 这可以从最简单的Burgers方程看到【19, 20】。因此，人们不得不面对复杂的情形，深入思考为什么应用Gauss-Green定理理解Cauchy通量非常困难。

任何流动都有一个动力学过程(dynamic process)，这也许是个常识，但需要严格论证。最近的研究表明【3, 4】，流动的时空关系既是相互独立又是彼此关联。下面的原理深刻地反映这一事实。

**有限体积方法基本原理.** 设$\boldsymbol{u}(\mathbf{x},\mathrm{t})$是定义1.1意义下方程(8)的弱解，$\Omega$是$\mathbb{R}^{\mathrm{n}}$中任一紧集。假定

(i) $\mathbf{u}(\mathbf{x},\mathrm{t})$是局部有界;

(ii) 质量$\mathbf{m}(\mathrm{t})$是时间t的连续函数。

那么有下列结论:

(i) 对任意$\delta>0$, 函数$\boldsymbol{H}(\boldsymbol{x};t,t+\delta t)=\int_{\mathrm{t}}^{t+\delta\mathrm{t}}\boldsymbol{F}(\boldsymbol{x},\mathrm{t})\,\mathrm{dt}$ 满足

$$\nabla_{\boldsymbol{x}}\cdot\boldsymbol{H}(\boldsymbol{x};t,t+\delta t)\in L_{loc}^{\infty}(\mathbb{R}^{n}). \tag{10}$$

(ii) 对任意$\delta>0$，时间段$(t,t+\delta t)$内流过$\Omega$边界$\partial\Omega$的总通量(lump flux)

$$\boldsymbol{\mathcal{F}}(\boldsymbol{\partial\Omega};t,t+\delta t)=\int_{t}^{t+\delta t}\int_{\boldsymbol{\partial\Omega}}\boldsymbol{F}\cdot\boldsymbol{n}\,\boldsymbol{d\Gamma dt} \tag{11}$$

关于$\Omega$的边界$\partial\Omega$扰动是Lipschitz连续的。

总通量$\boldsymbol{\mathcal{F}}(\boldsymbol{\partial\Omega};t,t+\delta t)$和Cauchy通量$\boldsymbol{\mathcal{C}}(\boldsymbol{\partial\Omega};\ \boldsymbol{t})$明显不同，特别是Lipschitz 连续性，进一步反映了流场的时空关联性质。而Cauchy通量只是一个瞬时行为，不能完全反映动力学的过程。这些反映了它们数学性质的根本差异。直观地来说，这里通过时间效应，换取空间的正则性。

由此给出(8)的弱解另一种表述，它恰恰是物理上常见的积分型平衡律。

**定义1.2.** 设$\Omega$是$\mathbb{R}^{n}$中任一紧集, $\delta t>0$. 如果它满足下列条件:

(i) $\boldsymbol{u}(\boldsymbol{x},t)$局部有界，且质量$\boldsymbol{m}(t)$是时间$t$的连续函数;

(ii) 整体通量(4)有意义且关于$\Omega$的边界$\partial\Omega$扰动是连续的;

(iii) 积分平衡律成立

$$\int_{\Omega}\boldsymbol{u}(\boldsymbol{x},t+\delta t)d\boldsymbol{x}-\int_{\Omega}\boldsymbol{u}(\boldsymbol{x},\boldsymbol{t})\boldsymbol{dx}=-\int_{t}^{t+\delta t}\int_{\partial\Omega}\boldsymbol{F}\cdot\boldsymbol{n}\,d\Gamma\,dt, \tag{12}$$

其中$\boldsymbol{n}$是$\partial\Omega$的单位外法向, 则称函数$\boldsymbol{u}(\boldsymbol{x},t)$是偏微分方程(8)的弱解。

事实上，已经证明定义1.1和1.2是等价的， 见【3,4】，并且给出了通量(11)的正则性质。这个定义说明将要讨论的有限体积方法就是计算流体力学的基本格式，和物理最原始的建模一致。流体力学方程组(8)可理解为: 如果流场光滑，用偏微分方程组(8)刻画；但当流场失去正则性时，解需要满足积分平衡律(12)。流体力学有限体积格式的构造过程就是基于(12)的直接建模过程：在偏微分方程(8)诱导出的解辅助下，构造出和(12)相容的离散模型，精准反映真实的物理过程。

## 三 有限体积方法及其时空耦合性

前面论述到：有限体积方法构造过程就是在偏微分方程(8)辅助下的一个直接建模过程。通常的有限体积方法从偏微分方程组(8)出发，在每一个计算单元（控制体）上应用 Gauss-Green 定理，得到有限体积公式。具体地，将计算区域$\Omega$剖分为$N$个计算单

元（控制体）$\Omega_j$，$\Omega = \bigcup_{j=1}^{N} \Omega_j$，$\partial\Omega_{\mathrm{j}} = \bigcup_{\ell} \Gamma_{j\ell}$. 有下列两种形式：半离散和全离散有限体积方法。

### 3.1. 半离散有限体积方法

在$\Omega_{\mathrm{j}}$上，对(8)的空间散度项应用 Gauss-Green 公式可得半离散有限体积方程

$$\frac{d}{dt}\int_{\Omega_{\mathrm{j}}} \boldsymbol{u}(\boldsymbol{x},t)\, d\boldsymbol{x} = -\int_{\partial\Omega_{\mathrm{j}}} \boldsymbol{F}\cdot\boldsymbol{n}\, d\Gamma, \quad j = 1,\cdots,N. \tag{13}$$

它直接对应于流体力学方程组(5)。在初始时刻$t = 0$, 给定$\boldsymbol{u}(\boldsymbol{x},t)$的分布，

$$\boldsymbol{u}(\boldsymbol{x},0) = \boldsymbol{P}_{\boldsymbol{0}}^{\boldsymbol{k}}(x) \in \mathcal{V}^k, \tag{14}$$

其中$\mathcal{V}^k$是一个可持续函数空间(通常为分片多项式函数空间[6])，意味着在每个时间层$t = t_n > 0$，$\boldsymbol{u}(\boldsymbol{x},t_n)$ 经过投影后所得$P_n^k \in \mathcal{V}^k$中。由于守恒性的基本要求

$$\int_{\Omega_j} \boldsymbol{u}(\boldsymbol{x},\boldsymbol{t_n})\, d\boldsymbol{x} = \int_{\Omega_j} \boldsymbol{P}_{\boldsymbol{n}}^{\boldsymbol{k}}(\boldsymbol{x}) \boldsymbol{dx}, \tag{15}$$

$\boldsymbol{P}_{\boldsymbol{n}}^{\boldsymbol{k}}(x)$一般是沿着每个控制体$\Omega_{\mathrm{j}}$的边界$\Gamma_{j\ell}$是间断的。为了利用(13)更新解，需要在每个控制体$\Omega_{\mathrm{j}}$的边界$\Gamma_{j\ell}$附近求解(8)，数值上使用后面所描述的黎曼解法器或近似黎曼解法器，得到在时刻$t$的数值通量

$$\left|\boldsymbol{\Gamma}_{\boldsymbol{j\ell}}\right| \boldsymbol{F}_{\boldsymbol{j\ell}}(t)\cdot \boldsymbol{n}_{\boldsymbol{j\ell}} \approx \int_{\Gamma_{\mathrm{j}\ell}} \boldsymbol{F}\cdot\boldsymbol{n}\, \boldsymbol{d\Gamma}, \tag{16}$$

带入(13)即得

$$\frac{d}{dt}\int_{\Omega_{\mathrm{j}}} \boldsymbol{u}(\boldsymbol{x},t)\, d\boldsymbol{x} = -\int_{\partial\Omega_{\mathrm{j}}} \boldsymbol{F}\cdot\boldsymbol{n}\, d\Gamma \approx -\sum_{\ell} \left|\boldsymbol{\Gamma}_{\boldsymbol{j\ell}}\right| \boldsymbol{F}_{\boldsymbol{j\ell}}(t)\cdot\boldsymbol{n}_{\boldsymbol{j\ell}}, \quad j = 1,\cdots,N. \tag{17}$$

接着使用常微分方程求解器，如典型的 Runge-Kutta 方法，求解这个时间依赖(常微分)方程。需要指出的是：这个方程需要在每个时间层$t = t_n$求解数值通量，并把(13)在下一时间层的解投影到$\mathcal{V}^k$中，得到$P_{n+1}^k \in \mathcal{V}^k$。如果用每个时间步用多级 Runge-Kutta 方法，则需要多次求解数值通量和多次投影。这个投影过程又称为数据重构。用算子形式，半离散方法可以表示为：

$$\boldsymbol{P}_{n+1}^{k}(\boldsymbol{x}) = \mathcal{P}\circ\mathcal{RK}\circ\mathcal{A}_{rie}\left[\boldsymbol{P}_{n}^{k}(\boldsymbol{x})\right], \tag{18}$$

其中$\mathcal{A}_{rie}$表示用黎曼解法器构造瞬时数值通量，$\mathcal{RK}$代表 Runge-Kutta 型的时间推进过程，$\mathcal{P}$表示数据重构。

[6] 用高阶多项式等光滑函数去逼近正则性很弱的函数(如间断函数)，常常会导致振荡现象(Gibbs 现象). 数据重构(数据投影)仍是有限体积格式中未彻底解决的难题。

我们注意到，由于对 Cauchy 通量$\int_{\Gamma_{j\ell}} \boldsymbol{F}\cdot\boldsymbol{n}\, \boldsymbol{d\Gamma}$还没有清晰的认识，不得不借助局部黎曼解法器(见下节)进行数值通量(16)的逼近。一般来说，所得到的局部误差为

$$\left|\boldsymbol{\Gamma}_{j\ell}\right| \boldsymbol{F}_{j\ell}(t)\cdot\boldsymbol{n}_{j\ell} - \int_{\Gamma_{j\ell}} \boldsymbol{F}\cdot\boldsymbol{n}\, \boldsymbol{d\Gamma} = \boldsymbol{\mathcal{O}}(|(\boldsymbol{\Delta u})_{jl}|)\left|\boldsymbol{\Gamma}_{j\ell}\right|, \tag{19}$$

这里$|(\boldsymbol{\Delta u})_{jl}|$表示数据$\boldsymbol{P}_n^k(\boldsymbol{x})$跨过控制体边界$\Gamma_{jl}$的局部跳跃。

半离散有限体积方法属于线方法(Method of Line)，从某种意义上来说是时空解耦方法。在光滑流场中，流体力学方程组的 PDE 关系可以直接反映时空关联性；但对间断解来说，(19)中的局部误差的累积给数值模拟带来巨大困难。

### 3.2. 全离散有限体积方法

本文将把(12)应用到时空控制体$\Omega_j \times [t_n, t_{n+1}]$, $t_{n+1} = t_n + \Delta t_n$, $\Delta t_n > 0$为时间步长，得到有限体积格式的全离散形式

$$\int_{\Omega_j} \boldsymbol{u}(\boldsymbol{x}, t_{n+1}) d\boldsymbol{x} - \int_{\Omega_j} \boldsymbol{u}(\boldsymbol{x}, \boldsymbol{t_n}) \boldsymbol{dx} = -\int_{t_n}^{t_{n+1}} \int_{\partial\Omega_j} \boldsymbol{F}\cdot\boldsymbol{n}\, d\Gamma\, dt. \tag{20}$$

也可以把(20)看作是(13)在时间段$(t_n, t_{n+1})$上的积分。与半离散情形下求解瞬时通量(16)不同，这里需要利用偏微分方程(8)的时空关联性质逼近$[t_n, t_{n+1}]$上总通量,

$$\sum_{\ell} \left|\boldsymbol{\Gamma}_{j\ell}\right| \boldsymbol{F}_{j\ell}(t_n, t_{n+1})\cdot\mathbf{n}_{j\ell} \approx \sum_{\ell} \int_{t_n}^{t_{n+1}} \int_{\Gamma_{j\ell}} \boldsymbol{F}\cdot\boldsymbol{n}\, \boldsymbol{d\Gamma dt}. \tag{21}$$

将之代入(20)，得到有限体积的公式

$$\overline{\boldsymbol{u}}_j^{n+1} = \overline{\boldsymbol{u}}_j^n - \frac{1}{|\Omega_j|} \sum_{\ell} \left|\boldsymbol{\Gamma}_{j\ell}\right| \boldsymbol{F}_{j\ell}(t_n, t_{n+1})\cdot\boldsymbol{n}_{jl}\,, \tag{22}$$

这里记逼近解在控制体$\Omega_j$上的平均值为

$$\bar{u}_j^n = \frac{1}{|\Omega_j|} \int_{\Omega_j} \boldsymbol{u}(\boldsymbol{x}, t_n) d\boldsymbol{x}. \tag{23}$$

由有限体积方法基本原理，可以得到

$$\left|\boldsymbol{\Gamma}_{j\ell}\right| \boldsymbol{F}_{j\ell}(t_n, t_{n+1}) - \int_{t_n}^{t_{n+1}} \int_{\Gamma_{j\ell}} \boldsymbol{F}\cdot\boldsymbol{n}\, \boldsymbol{d\Gamma dt} = \boldsymbol{\mathcal{O}}(\boldsymbol{\Delta t_n})^{q+1} \left|\boldsymbol{\Gamma}_{j\ell}\right|, \tag{24}$$

其中$q > 0$。值得注意的是，(24)中的误差和(19)中的误差非常不同：(19)中的误差是用解的局部跳跃（变差）来测量，而(24)中的误差直接用时间步长(等价于网格尺度）来测量。当流场相对光滑时，这两者是等价的；但当流场中有剧烈间断、脉动时，二

者差别是明显的. 即使网格加密，(19)也得不到收敛解[7]【7】。后文算例 6.1 可以看出数值通量的重要性。

同半离散方法相似，在每个时间层$t = t_n$上投影可得到在可持续空间$\mathcal{V}^k$中的逼近解$P_n^k$。全离散有限体积方法可以用符号表示为：

$$\boldsymbol{P}_{n+1}^k(\boldsymbol{x}) = \mathcal{P} \circ \mathcal{E} \circ \mathcal{A}[\boldsymbol{P}_n^k(\boldsymbol{x})], \tag{25}$$

这里$\mathcal{A}$表示通量的逼近，$\mathcal{E}$表示解的有限体积发展过程，$\mathcal{P}$代表投影过程(数据重构)。一旦$\mathcal{A}$被确定，$\mathcal{E}$是自然的且没有任何误差产生。数据重构部分是重要的，将在后面评注。

### 3.3. 有限体积方法的相容性

所谓相容性，描述的是离散格式与背景方程之间的关系。传统上常常使用 Taylor 展开的方式研究数值格式与微分方程(如(8))之间的相容性，这样的运算只对光滑流场是成立的。当流场比较复杂时，目前大部分的数值分析某种意义上只是启发性的，比如以标量方程为模型建立相关的理论【21】。

现在基于积分平衡律(12)来建立有限体积格式的相容性，即离散格式(18)或(25)须与(12)进行直接比较。由(18)或(25)可知, 误差来源于数据投影步$\mathcal{P}$以及通量的逼近步$\mathcal{A}_{rie}$或$\mathcal{A}$。对于前者，虽然有大量的文献存在，数据重构步主要依赖相关的空间信息。文【22】只是做了初步的尝试，把未来时间的数据重构与过去的信息进行了关联。本文假定投影步满足所有的精度要求，只讨论通量的逼近误差。

对半离散格式(18)来说，在每个固定时刻给出的通量误差至多为

$$\sum_{\ell} |\boldsymbol{\Gamma}_{j\ell}| \ \ \boldsymbol{F}_{j\ell}(t) \cdot \boldsymbol{n}_{j\ell} - \sum_{\ell} \int_{\Gamma_{j\ell}} \boldsymbol{F} \cdot \boldsymbol{n} \, \boldsymbol{d\Gamma} = \boldsymbol{\mathcal{O}}\big((\boldsymbol{\Delta u})_{jl}\big) |\boldsymbol{\Gamma}_{j\ell}| = \mathcal{O}\big((\boldsymbol{\Delta u})_j\big) |\boldsymbol{\Gamma}_j| d_j, \tag{26}$$

其中$(\boldsymbol{\Delta u})_j$表示在控制体附近解的局部跳跃，$|\boldsymbol{\Gamma}_j|$是 $\Omega_j$边的最大长度，$d_j = diam(\Omega_j)$。并且在每个时间层，随着 Runge-Kutta 步的增加，误差也会累积。在(26)中，$d_j$来源于不同方向上通量差的获利，在 4.5 节可以进一步看到。

对于全离散方法(25)，我们讨论每个时间层的数值通量的逼近，并期望有下列的估计

$$\sum_{\ell} |\boldsymbol{\Gamma}_{j\ell}| \ \boldsymbol{F}_{j\ell}(t_n, t_{n+1}) - \sum_{\ell} \int_{t_n}^{t_{n+1}} \int_{\Gamma_{j\ell}} \boldsymbol{F} \cdot \boldsymbol{n} \, \boldsymbol{d\Gamma dt} = \boldsymbol{\mathcal{O}}(\boldsymbol{\Delta t_n})^{q+2} |\boldsymbol{\Gamma}_j| \, . \tag{27}$$

其中$q > 0$。比较(26)和(27)，它们有本质的差别，即$(\boldsymbol{\Delta u})_j$与$\boldsymbol{\Delta t_n}$的差别。对光滑解来说它们是等价的。这样我们给出下列的定义：

---

[7]在标量方程的研究中，网格加密是可以得到收敛解的，原因在于全局变差有界性质. 到目前为止还没有例证可证明该性质在流体力学方程组成立。

**定义 3.1（有限体积格式的相容性）**. 设$\Omega_j$是任一控制体，$j = 1,\cdots,N$，如果对于某个$q > 0$，(27)成立，则有限体积格式(25)相容于平衡律(12)，并具有$q$阶相容性。

如上所述，相容性概念常常相对与微分方程所言. 如果(8)是双曲守恒律，Lax 和 Wendroff 给出了有限差分方法的相容性，常称为 Lax 相容性【23】。由于 Lax 相容性概念影响深远，有必要进行回顾。

**定义 3.2（Lax 相容性【23】）**。考虑双曲守恒律方程组

$$\partial_t \boldsymbol{u} + \partial_x \boldsymbol{f}(\boldsymbol{u}) = 0, \tag{28}$$

其$2p+1$点守恒性差分格式为

$$\boldsymbol{u}_j^{n+1} = \boldsymbol{u}_j^n - \lambda \left[\boldsymbol{g}_{j+\frac{1}{2}}^n - \boldsymbol{g}_{j-\frac{1}{2}}^n\right], \qquad \lambda = \frac{\Delta t}{\Delta x}, \tag{29}$$

其中$\Delta x$是网格尺寸，$\Delta t$是时间步长，$\boldsymbol{g}_{j+\frac{1}{2}}^n = \boldsymbol{g}(\boldsymbol{u}_{j-p+1}^n,\cdots,\boldsymbol{u}_{j+p}^n)$。如果成立

$$\boldsymbol{g}(\boldsymbol{u},\cdots,\boldsymbol{u}) = \boldsymbol{f}(\boldsymbol{u}), \tag{30}$$

则称差分方法（29）与双曲守恒律(28)相容。

基于这个定义，建立了影响深远的 Lax-Wendroff 收敛定理，直至今日，该定理仍被广泛应用。但是随着高阶精度格式的发展以及工程应用的需求，这个定理常常被误用，典型表现为:

(i) 相容性定义(30)只适用于传统一阶精度数值方法，对于目前广泛使用的高阶精度方法并不使用，特别是高阶有限体积方法等。

(ii) 在此基础上建立的 Lax-Wendroff 定理只适用于一致网格或结构网格，对非结构网格并不适用【24,25】. 数值验证过程中使用的网格加密方法测试收敛阶，需要倍加小心。

定义 3.1 实际上第一次给出高精度有限体积数值格式与积分平衡律之间的相容性【3】。有限体积方法与网格的结构无关，因此适用于无结构网格。特别地，收敛阶测试时一般针对光滑流进行，数据重构部分能够达到应有的精度，通量逼近阶事实上就是收敛阶。但是，当流场含有间断或其它复杂结构时，数据重构仍存在诸多争议，通量的逼近效果往往被忽略。通过以上分析可以看到，如果通量相容性(27)不成立，逼近解不可能收敛。也就是说，（27）是有限体积格式相容性的一个必要条件。

### 3.4. 再访 Godunov 方法

谈到流体力学中的有限体积方法，需要回顾 Godunov 方法【26】。 经过半个多世纪的发展和检验，它已成为现代计算流体力学基石。考虑一维可压缩欧拉方程组

$$\partial_t \boldsymbol{u} + \partial_x \boldsymbol{F}(\boldsymbol{u}) = 0, \tag{31}$$

其中守恒量$\boldsymbol{u}=(\rho,\rho u,\rho E)^{\top}$，$\boldsymbol{F}(\boldsymbol{u})=\left(\rho u,\rho u^2+p,u(\rho E+p)\right)^{\top}$. 这个方程组由 Gibbs 关系式封闭，这里不再赘述。记时空控制体$C_j^n=I_j\times[t_n,t_{n+1}]$，$I_j=(x_{i-\frac{1}{2}},x_{j+\frac{1}{2}})$，$x_j=\frac{1}{2}\left(x_{j-\frac{1}{2}}+x_{j+\frac{1}{2}}\right)$, $\Delta x_j=x_{j+\frac{1}{2}}-x_{j-\frac{1}{2}}$，$t_{n+1}=t_n+\Delta t_n$。 那么(31)可以写成

$$\int_{I_j}\boldsymbol{u}(x,t_{n+1})dx-\int_{I_j}\boldsymbol{u}(x,t_n)dx+\int_{t_n}^{t_{n+1}}\boldsymbol{F}\left(\boldsymbol{u}\left(x_{j+\frac{1}{2}},t\right)\right)dt-\int_{t_n}^{t_{n+1}}\boldsymbol{F}\left(\boldsymbol{u}\left(x_{j-\frac{1}{2}},t\right)\right)dt=0, \tag{32}$$

即积分平衡律(12)的形式。对应于(22)有限体积格式为

$$\overline{\boldsymbol{u}}_j^{n+1}=\overline{\boldsymbol{u}}_j^{n}-\frac{\Delta t_n}{\Delta x_j}\left[\boldsymbol{F}_{j+\frac{1}{2}}^{\boldsymbol{n}}-\boldsymbol{F}_{j-\frac{1}{2}}^{\boldsymbol{n}}\right], \tag{33}$$

其中$\boldsymbol{F}_{j+\frac{1}{2}}^{\boldsymbol{n}}$是数值通量。给定初始的数据分布$\boldsymbol{u}(\boldsymbol{x},\boldsymbol{t_n})=\boldsymbol{P_n^0}\in\boldsymbol{\mathcal{V}^0}$, Godunov 格式的一个中心思想是在每个网格边界点$\left(x_{j+\frac{1}{2}},t_n\right)$处求解相应黎曼问题

$$\begin{aligned}&\partial_t\boldsymbol{u}+\partial_x\boldsymbol{F}(\boldsymbol{u})=0,\qquad x\in\mathbb{R},\ t>t_n,\\ &\boldsymbol{u}(x,t_n)=\begin{cases}\overline{\boldsymbol{u}}_j^n, & x<x_{j+1/2},\\ \overline{\boldsymbol{u}}_{j+1}^n, & x>x_{j+1/2}.\end{cases}\end{aligned} \tag{34}$$

求得网格边界上解的值$\boldsymbol{u}_{j+\frac{1}{2}}^{\boldsymbol{n}}$。欧拉方程组(31)的黎曼解可以清晰给出，记为$\boldsymbol{u}(x,t)=R(\xi,\boldsymbol{u_j^n},\boldsymbol{u_{j+1}^n})$，$\xi=\frac{x-x_{j+\frac{1}{2}}}{t-t_n}$，$\boldsymbol{u}_{j+\frac{1}{2}}^{\boldsymbol{n}}=\boldsymbol{R}\,(\boldsymbol{0};\overline{\boldsymbol{u}}_{\boldsymbol{j}}^{\boldsymbol{n}},\overline{\boldsymbol{u}}_{\boldsymbol{j+1}}^{\boldsymbol{n}})$。 这样$\boldsymbol{F}_{j+\frac{1}{2}}^{\boldsymbol{n}}=\boldsymbol{F}(\boldsymbol{u}_{j+\frac{1}{2}}^{n})$， 从而得到 Godunov 格式，

$$\overline{\boldsymbol{u}}_j^{n+1}=\overline{\boldsymbol{u}}_j^{n}-\frac{\Delta t_n}{\Delta x_j}\left[\boldsymbol{F}\left(\boldsymbol{u}_{j+\frac{1}{2}}^{n}\right)-\boldsymbol{F}\left(\boldsymbol{u}_{j-\frac{1}{2}}^{n}\right)\right]. \tag{35}$$

这是和(12)完全一致的

$$\frac{1}{\Delta t_n}\int_{t_n}^{t_{n+1}}\boldsymbol{F}\left(\boldsymbol{u}\left(x_{j+\frac{1}{2}},t\right)\right)dt=\boldsymbol{F}\left(\boldsymbol{u}_{j+\frac{1}{2}}^{n}\right).$$

因此，从积分平衡律的逼近来说，基于分片常数的初值逼近，(35)与(12)完全相容或有无穷阶相容性。从整体逼近(25)来说，所有的误差来自于投影运算， 因此习惯上称 Godunov 格式为一阶格式。

如果用逼近的黎曼解法解，界面上的值一般可以写为

$$\widetilde{\boldsymbol{u}}^{n}{}_{j+\frac{1}{2}}=\frac{1}{2}\left(\overline{\boldsymbol{u}}_j^n+\overline{\boldsymbol{u}}_{j+1}^n\right)-\frac{\alpha\left(\overline{\boldsymbol{u}}_j^n,\overline{\boldsymbol{u}}_{j+1}^n\right)}{2}\left[\boldsymbol{F}\left(\overline{\boldsymbol{u}}_{j+1}^n\right)-\boldsymbol{F}\left(\overline{\boldsymbol{u}}_j^n\right)\right], \tag{36}$$

这里$\alpha\left(\overline{\boldsymbol{u}}_j^n,\overline{\boldsymbol{u}}^{n}{}_{j+1}\right)$称为粘性系数。一般来说，通量的逼近误差为

$$\frac{1}{\Delta t_n}\int_{t_n}^{t_{n+1}}\boldsymbol{F}\left(\boldsymbol{u}\left(x_{j+\frac{1}{2}},t\right)\right)dt-\boldsymbol{F}\left(\widetilde{\boldsymbol{u}}^{n}{}_{j+\frac{1}{2}}\right)=\mathcal{O}(|\overline{\boldsymbol{u}}_{j+1}^n-\overline{\boldsymbol{u}}_j^n|). \tag{37}$$

这个误差在强间断的情形下对计算结果有巨大的伤害。如(19)所述，这个误差不随着网格加密而减小。

在多维情形下，例如二维双曲守恒律情形

$$\partial_t \boldsymbol{u} + \partial_x F(\boldsymbol{u}) + \partial_y \boldsymbol{G}(\boldsymbol{u}) = 0, \tag{38}$$

在每个时间层$t = t_n$的数据为分片常数，即每个控制体$\boldsymbol{\Omega}_j$内为常数$\bar{\boldsymbol{u}}_j^n$, 通过求解相应的黎曼问题, 构造数值通量。这时相应的黎曼问题可分为两类：

（i） 相对于界面的法向黎曼问题

由于流体力学方程组的伽利略(Galilean)不变性，通过旋转变换，总可以设$x$-方向为$\Gamma_{j\ell}$的外法向，$\Gamma_{j\ell}$两侧单元的值记为$\boldsymbol{u}_L$和$\boldsymbol{u}_R$，法向黎曼问题定义为

$$\begin{aligned} &\boldsymbol{\partial_t u} + \partial_x \boldsymbol{F}(\boldsymbol{u}) = 0, \quad (x, y) \in \mathbb{R}^2, \quad t > 0, \\ &\boldsymbol{u}(x, y, 0) = \begin{cases} \boldsymbol{u}_L, & x < 0, \\ \boldsymbol{u}_R, & x > 0. \end{cases} \end{aligned} \tag{39}$$

它的求解和上面的一维黎曼问题(34)没有本质差异。显然，此解只反映了沿$\Gamma_{j\ell}$法向流场的变化情况。

（ii） 顶点处二维黎曼问题

这是真正的多维问题，可以写成如下形式【27】

$$\begin{aligned} &\partial_t \boldsymbol{u} + \partial_x F(\boldsymbol{u}) + \partial_y \boldsymbol{G}(\boldsymbol{u}) = 0, \quad (x, y) \in \mathbb{R}^2, \quad t > 0, \\ &\boldsymbol{u}(x, y, 0) = \boldsymbol{u}_\ell, \qquad (x, y) \in \Theta_\ell, \end{aligned} \tag{40}$$

这里$\Theta_\ell$，$\ell = 1, \cdots, K$, 是从原点出发的角形区域, 见图 3.1。由于有限传播性质，从中心点发出的复杂的波结构只会影响有限的区域。仅从通量的构造来说，顶点处解对界面通量的贡献占比很小, 可以适当限制 Courant 数，减少顶点周边流场对数值通量的影响。在移动网格方法中，特别是拉格朗日方法中，需要用顶点解确定顶点运动速度，顶点处二维黎曼问题的解非常重要。但由于解的结构过于复杂，往往通过顶点近似黎曼解法器来处理【28】。

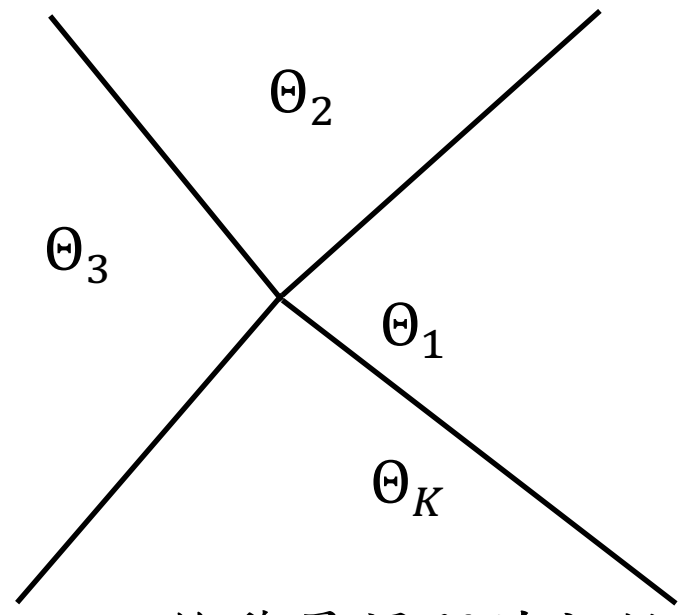

图 3.1. 二维黎曼问题的初始数据分布

黎曼问题的（逼近）解被广泛用到双曲守恒律及延展到更一般的流体力学方程组半离散数值方法中。与下面广义黎曼问题的解进行比较可以发现，这个解不能充分反映出流体的动力学过程；即使从微观角度，欧拉方程反映了粒子无穷碰撞的结果。再

者，在每个控制单元上及初始时间层，流场处于常状态，空间的变化依赖相邻单元之间的间断来实现。因此，Godunov 格式的解不能充分反映瞬时行为。对长时间、多物理以及多尺度现象，数值模拟结果常常出现似是而非的现象。

到目前为止黎曼问题只限于对双曲守恒律进行研究， 相关的拓展可以从下面的广义黎曼问题研究中看出。

### 3.5. 高阶数值通量与广义黎曼问题

高阶数值通量的构造实质上等价于相应广义黎曼问题的数值求解，即广义黎曼问题解法器以及高阶数值实现(算法构造)。即在$t = t_n$时刻，给定初始数据 $P_n^k \in \mathcal{V}^k$，求解方程组(8)。与黎曼问题类似， 广义黎曼问题包括界面法向广义黎曼问题以及顶点广义黎曼问题。后者只是在依赖网格顶点速度的移动网格方法中起着重要作用，有兴趣的读者可参阅【28】。下面只讨论前者。

一般地，方程组(8)广义黎曼问题可以写成如下形式

$$\begin{aligned} &\partial_t \boldsymbol{u} + \nabla \cdot \boldsymbol{F}(\boldsymbol{u}, \nabla \boldsymbol{u}, \cdots) = 0, \quad \boldsymbol{x} \in \mathbb{R}^n, t > 0, \\ &\boldsymbol{u}(\boldsymbol{x}, \boldsymbol{t} = \boldsymbol{0}) = \begin{cases} \boldsymbol{u_L}(\boldsymbol{x}), & \boldsymbol{\Phi}(\boldsymbol{x}) < \boldsymbol{0}, \\ \boldsymbol{u_R}(\boldsymbol{x}), & \boldsymbol{\Phi}(\boldsymbol{x}) > \boldsymbol{0}, \end{cases} \end{aligned} \tag{41}$$

其中$\boldsymbol{\Phi}(\boldsymbol{x}) = \boldsymbol{0}$是定向的$n-1$维光滑曲面，经过适当的坐标变换，可记该曲面为$x = 0$ (记空间坐标为$\boldsymbol{x} = (x, y, z)$)，即(41)可化为如下形式，

$$\begin{aligned} &\partial_t \boldsymbol{u} + \nabla \cdot \widetilde{\boldsymbol{F}}(\boldsymbol{u}, \nabla \boldsymbol{u}, \cdots) = \widetilde{\boldsymbol{H}}(\boldsymbol{x}, \boldsymbol{u}), \quad \boldsymbol{x} \in \mathbb{R}^n, t > 0, \\ &\boldsymbol{u}(\boldsymbol{x}, \boldsymbol{t} = \boldsymbol{0}) = \begin{cases} \boldsymbol{u_L}(\boldsymbol{x}), & x < \boldsymbol{0}, \\ \boldsymbol{u_R}(\boldsymbol{x}), & x > \boldsymbol{0}, \end{cases} \end{aligned} \tag{42}$$

其中$x = 0$称为界面. 这里未知量仍采用同样的记号，由变换后所得形式$\widetilde{\boldsymbol{H}}(\boldsymbol{x}, \boldsymbol{u})$来自于曲面$\boldsymbol{\Phi}(\boldsymbol{x}) = \boldsymbol{0}$的几何效果，下面讨论中将省略记号“～”。可以看出，广义黎曼问题与相应黎曼问题至少有三方面不同：(i) 初始数据不同，$\boldsymbol{u_L}(\boldsymbol{x})$与$\boldsymbol{u_R}(\boldsymbol{x})$是非常状态的两个光滑函数. 除了精度以外，数据的非一致性包含更多的物理信息，比如熵变化。(ii) 黎曼问题只对双曲守恒律而言，而广义黎曼问题的控制方程可以任意的偏微分方程，比如具有粘性力、外力等效应的流体力学方程组，或更一般的 Boltzmann 型方程。(iii) 正如在(41)所表达那样，当$\boldsymbol{\Phi}(\boldsymbol{x}) = \boldsymbol{0}$不是一个平面时，它的拓扑变化也嵌入到解的变化中。

根据有限体积方法基本原理，需要直接逼近

$$\boldsymbol{\mathcal{F}}(A_\epsilon; 0, \delta) = \int_{A_\epsilon} \int_0^\delta \boldsymbol{F}\big(\boldsymbol{u}(0, y, z; t)\big) dydz\, dt, \tag{43}$$

这里$A_\epsilon = \{(0, y, z); y \in (y_0 - \epsilon, y_0 + \epsilon), z \in (z_0 - \epsilon, z_0 + \epsilon)\}$是面$x = 0$上的一部分。在一维情况下，(43) 即为

$$\boldsymbol{\mathcal{F}}(0; 0, \delta t) = \int_0^{\delta t} \boldsymbol{F}\big(\boldsymbol{u}(0; t)\big) dt. \tag{44}$$

根据已有流体力学方程组相关的偏微分方程理论与应用研究，可以有效地逼近或求解(42)，满足与(27)对应的相容性要求。具体地，(44)有

$$\boldsymbol{\mathcal{F}}(0;0,\delta t)=\delta t\left[\boldsymbol{F}\big(\boldsymbol{u}(0;0^+)\big)+\frac{\delta t}{2}\partial_t\boldsymbol{F}\big(\boldsymbol{u}(0;0^+)\big)\right]+\mathcal{O}(\delta t^3). \tag{45}$$

这里需要被积函数$\boldsymbol{F}\big(\boldsymbol{u}(0;t)\big)$满足关于时间$t$的正则性. 对于(42)中的初始数据，这一点可以得到保障，从而可以对(44)进行相容性逼近。

**定义3.3（有限体积方法的时空耦合性）.** 考虑时空关联模型(8)的有限体积方法(25). 如果方程(8)的解可以用来逼近数值通量，并使(25)在定义3.1的意义下相容，数据重构也充分利用(8)的时空关联信息，则算法(25)是时空耦合的。

**注3.4.** 文【22】给出的Hermite型数据重构方法是时空耦合的，该方法已用在两步四阶方法【4,29】中。文献中数据重构的时空耦合性研究还不充分，上述定义中“数据重构也充分利用(8)的时空关联信息”这句话比较模糊，需要做更深入的研究。

### 3.6. 几个时空耦合算法的例子

下面给出几个时空耦合算法的例子。

（i）线性输运方程

对于第一节中引出的线性方程(1)，其有限体积格式可以写成

$$\bar{u}_j^{n+1}=\bar{u}_j^n-\lambda_j^n\Big[\frac{1}{\Delta \mathrm{t}}\int_{t_n}^{t_{n+1}}u\left(x_{j+\frac{1}{2}},t\right)dt-\frac{1}{\Delta \mathrm{t}}\int_{t_n}^{t_{n+1}}u\left(x_{j-\frac{1}{2}},t\right)dt\Big],\quad \lambda=\frac{a\Delta t_n}{\Delta x_j}. \tag{46}$$

假设$a>0$，$\lambda_j^n<1$。如果初始数据是分片常数

$$u(x,t_n)=\bar{u}_j^n,\qquad\qquad x\in I_j, \tag{47}$$

则解在$(t_n,t_{n+1})$时间内有$u\left(x_{j+\frac{1}{2}},t\right)=\bar{u}_j^n$，及

$$\frac{1}{\Delta \mathrm{t}}\int_{t_n}^{t_{n+1}}u\left(x_{j+\frac{1}{2}},t\right)dt=\bar{u}_j^n. \tag{48}$$

从而由(46)可得

$$\bar{u}_j^{n+1}=\bar{u}_j^n-\lambda_j^n\big(\bar{u}_j^n-\bar{u}_{j-1}^n\big). \tag{49}$$

这就是迎风格式。数值通量的构造完全使用了解的信息，因此迎风格式(49)是时空耦合的。众所周知，迎风格式在所有一阶稳定格式中具有“最优”性质。

如果初始数据是分片多项式，特别是分片线性函数时

$$u(x,t_n)=\bar{u}_j^n+\sigma_j^n(x-x_j),\quad x\in I_j, \tag{50}$$

则(1)的解(2)可写为

$$u\left(x_{j+\frac{1}{2}},t\right)=\bar{u}_j^n+\sigma_j^n\left(\frac{\Delta x}{2}-a(t-t_n)\right),\ t\in(t_n,t_{n+1}). \tag{51}$$

直接计算可得

$$\frac{1}{\Delta t_n}\int_{t_n}^{t_{n+1}}u\left(x_{j+\frac{1}{2}},t\right)dt=u\left(x_{j+\frac{1}{2}},t_n+\frac{\Delta t}{2}\right)=:u_{j+\frac{1}{2}}^{n+\frac{1}{2}}, \tag{52}$$

以及

$$\begin{cases}\bar{u}_j^{n+1}=\bar{u}_j^n-\lambda_j^n\left(u_{j+\frac{1}{2}}^{n+\frac{1}{2}}-u_{j-\frac{1}{2}}^{n+\frac{1}{2}}\right),\ \lambda_j^n=\dfrac{a\Delta t_n}{\Delta x_j},\\[2ex] \sigma_j^{n+1}=\dfrac{1}{\Delta x_j}\left(u\left(x_{j+\frac{1}{2}},t_{n+1}\right)-u\left(x_{j-\frac{1}{2}},t_{n+1}\right)\right).\end{cases} \tag{53}$$

注意梯度更新$\sigma_j^{n+1}$也使用了解(51)的信息。这样构造的格式完全利用了解(51)的信息，所以格式(53)是时空耦合的。这里梯度$\sigma_j^{n+1}$更新后得到的分片线性函数仍可能有震荡现象，并不能有效逼近间断函数。从这一简单例子可以看出，数据重构仍然是值得探讨的。

更一般地，我们可以利用【29，22】的方式来构造高阶时空耦合方法，或者 Lax-Wendroff 型的单步高阶方法，只是推广到实际工程问题时，真正非线性和多维性会给单步方法带来实质性的困难。

（ii） 线性对流扩散方程

考虑下列方程

$$\partial_t u+a\partial_x u=\mu\partial_x^2 u,\qquad t>0, \tag{54}$$

这里物理粘性系数$\mu>0$. 把它写成散度形式有

$$\partial_t u+\partial_x(au-\mu\partial_x u)=0,\qquad \mu>0. \tag{55}$$

然后在时空控制体上，把它表示为(32)的形式，进行通量逼近。文【8】中的守恒元/解元(Conservation Element/Solution Element)方法展示其时空耦合的属性。由于其构造需要一点篇幅展示，有兴趣的读者可以查阅【8】及其后来的一系列文章。

相对来说，Navier-Stokes 方程组及其相关时空关联模型的时空耦合算法研究较少. 尽管形式上可以进行简单的时空对换处理，但对耗散项等高阶空间导数项的处理仍需要严格数学论证。当然，GKS 方法间接提供的 Navier-Stokes 方程的数值算法具有时空耦合性【9，30】。

（iii）线性松弛系统

考虑下列简单的松弛系统

$$\begin{aligned} &\partial_t u + a\partial_x u = \frac{v-u}{\tau} =: \frac{Q(u,v)}{\tau}, \qquad t>0, \\ &u(x,0) = u_0(x), \end{aligned} \tag{56}$$

其中 $\tau > 0$是松弛参数，$a$为常数，$v$也是常数。这个方程的有限体积形式为：

$$\bar{u}_j^{n+1} = \bar{u}_j^n - \lambda\left(u_{j+\frac{1}{2}}^{n+\frac{1}{2}} - u_{j-\frac{1}{2}}^{n+\frac{1}{2}}\right) + \frac{1}{\Delta x}\int_{t_n}^{t_{n+1}}\int_{x_{j-\frac{1}{2}}}^{x_{j+\frac{1}{2}}} \frac{Q(v,u)}{\tau}dxdt. \tag{57}$$

为了构造时空耦合算法，可以用(56)的解表达式，

$$u(x,t) = (u_0(x-at) - v)e^{-\frac{t}{\tau}} + v, \tag{58}$$

来计算数值积分$u_{j+\frac{1}{2}}^{n+\frac{1}{2}}$以及$\frac{Q(v,u)}{\tau}$的积分。为了更一般的应用，这里使用【30】中的方法，即 DUGKS 方法，得到

$$\bar{u}_j^{n+1} = \bar{u}_j^n - \lambda\left(u_{j+\frac{1}{2}}^{n+\frac{1}{2}} - u_{j-\frac{1}{2}}^{n+\frac{1}{2}}\right) + \frac{\Delta t}{2\tau}\left(Q_j^n + Q_j^{n+1}\right), \tag{59}$$

其中$u_{j+\frac{1}{2}}^{n+\frac{1}{2}}$是由下列形式给出

$$u_{j+\frac{1}{2}}^{n+\frac{1}{2}} = u\left(x_{j+\frac{1}{2}} - \frac{a}{2}\Delta t, t_n\right) + \frac{1}{2\tau}\left(Q\left(x_{j+\frac{1}{2}} - \frac{a}{2}\Delta t, t_n\right) + Q_{j+\frac{1}{2}}^{n+\frac{1}{2}}\right). \tag{60}$$

进而可以利用(58)或使用(60)类似的想法，得到$u\left(x_{j+\frac{1}{2}}, t_{n+1}\right)$，并用来更新梯度

$$\sigma_j^{n+1} = \frac{1}{\Delta x_j}\left(u\left(x_{j+\frac{1}{2}}, t_{n+1}\right) - u\left(x_{j-\frac{1}{2}}, t_{n+1}\right)\right). \tag{61}$$

很显然，格式(59)-(61)是时空耦合的，并具有时空二阶精度。

**四 基于广义黎曼解法器（GRP solver）的时空耦合算法**

数值通量的构造是执行有限体积格式的两个核心步骤之一。对于非线性问题我们一般无法像上述线性方程那样,得到解的表达式。下面将要利用广义黎曼问题(41)或(42)解的局部正则性质，发展广义黎曼解法器。其主要思想可以概括为：

| Lax-Wendroff 型解法器 + 奇异性分辨（间断追踪） |
| --- |

为了方便叙述，这节统一把界面放在$x=0$，把广义黎曼问题的求解点平移到坐标原点(0,0)，初始数据表示成(42)的形式。记

$$\boldsymbol{u}^*:=\lim_{t\to0^+}\boldsymbol{u}(\boldsymbol{0},t),\ \ (\partial_t\boldsymbol{u})^*:=\lim_{t\to0^+}\partial_t\boldsymbol{u}(\boldsymbol{0},t). \tag{62}$$

**广义黎曼问题（GRP）解法器：** 给定控制方程以及形如(42)的分片光滑函数作为初始数据，求解 (42) 并得到形如(61)的极限值。

一旦有了极限值(62)，受益于解$\boldsymbol{u}(\boldsymbol{x},t)$在界面上关于时间$t$的连续性，我们有

$$\boldsymbol{u}(0,\delta t)=\boldsymbol{u}^*+(\partial_t\boldsymbol{u})^*\delta t+\mathcal{O}(\delta t^2). \tag{63}$$

由于$\boldsymbol{u}^*$只表示了一种瞬时行为，其解由相应的黎曼解给定

$$\boldsymbol{u}^*=R\big(0;\boldsymbol{u}_{\boldsymbol{L}}(0),\boldsymbol{u}_{\boldsymbol{R}}(0)\big). \tag{64}$$

接下来的关键任务是求值$(\partial_t\boldsymbol{u})^*$。“非常不严格”地由控制方程（42）直接得到

$$\partial_t\boldsymbol{u}=-\nabla\cdot\widetilde{\boldsymbol{F}}(\boldsymbol{u},\nabla\boldsymbol{u},\cdots)+\widetilde{\boldsymbol{H}}(\boldsymbol{x},\boldsymbol{u}), \tag{65}$$

然后取极限。这样解$u(x,t)$的变化率可以通过空间的变化反映，就是经典的 Lax-Wendroff 方法【23】，或 Cauchy-Kowalevski 方法【31】。从中可以看出，通过这种途径把模型的时空关联性质嵌入到数值格式中，从而所得的算法具有时空耦合性质。

**注：**一般的单步 Lax–Wendroff 方法需要计算解的高阶时间导数$\partial_t^k\mathbf{u}$，但对流体力学方程组来说，需要避免这样的操作，原因是:（i）由于解中常有间断，这样的运算失去了数学与物理意义；（ii）即使允许如此运算，所得出的通量估算过于复杂，失去了实际工程意义；（iii）最近发展的两步四阶方法【4，29】表明，这一对值具有明确物理意义，可以用来作为构造高阶方法的基石，并避免高阶时间导数的运算。

### 4.1 线性 GRP 解法器

对于线性系统

$$\begin{aligned}&\boldsymbol{\partial_t u}+\boldsymbol{A\partial_x u}=\boldsymbol{Cu}+\boldsymbol{D},\quad \boldsymbol{t>0},\\ &\boldsymbol{u(x,0)}=\begin{cases}\boldsymbol{u_L(x)}, & \boldsymbol{x<0},\\ \boldsymbol{u_R(x)}, & \boldsymbol{x>0},\end{cases}\end{aligned} \tag{66}$$

其中$\boldsymbol{A},,\boldsymbol{C},\boldsymbol{D}$是常数矩阵，$\boldsymbol{A}$可以对角化，$\boldsymbol{\Lambda}=\boldsymbol{R}^{-1}\mathbf{A}\boldsymbol{R}$，其中$\boldsymbol{\Lambda}$是由$\boldsymbol{A}$的特征值$\lambda_i$组成的矩阵，$\boldsymbol{\Lambda}=diag\{\lambda_i\}$。记$\boldsymbol{w}=\boldsymbol{R}^{-1}\boldsymbol{u}$，有

$$\begin{aligned}&\boldsymbol{\partial_t w}+\boldsymbol{\Lambda\partial_x w}=\boldsymbol{R^{-1}Cu}+\boldsymbol{R^{-1}D},\quad \boldsymbol{t>0},\\ &\boldsymbol{w(x,0)}=\begin{cases}\boldsymbol{w_L(x)}, & \boldsymbol{x<0},\\ \boldsymbol{w_R(x)}, & \boldsymbol{x>0},\end{cases}\end{aligned} \tag{67}$$

进而记$\boldsymbol{\Lambda}^{+}=diag(\max(\lambda_i,0)\}$，$\boldsymbol{\Lambda}^{-}=diag(\min(\lambda_i,0)\}$，$\boldsymbol{I}^{+}=\frac{1}{2}diag\{1+sign(\lambda_i)\}$，$\boldsymbol{I}^{-}=-\frac{1}{2}diag\{1-sign(\lambda_i)\}$. 显然我们可到

$$(\partial_t \boldsymbol{w})^{*}=-(\boldsymbol{\Lambda}^{+}\boldsymbol{w}_L'+\boldsymbol{\Lambda}^{-}\boldsymbol{w}_R')+\left(\boldsymbol{I}^{+}\boldsymbol{R}^{-1}\boldsymbol{C}\boldsymbol{u}_L+\boldsymbol{I}^{-}\boldsymbol{R}^{-1}\boldsymbol{C}\boldsymbol{u}_R\right)+\boldsymbol{R}^{-1}\boldsymbol{D}, \tag{68}$$

其中$\boldsymbol{u}_L=\boldsymbol{u}_L(\boldsymbol{0})$，$\boldsymbol{u}_R=\boldsymbol{u}_R(\boldsymbol{0})$，$\boldsymbol{w}_L'=\boldsymbol{w}_L'(\boldsymbol{0})$，$\boldsymbol{w}_R'=\boldsymbol{w}_R'(\boldsymbol{0})$. 再记$\boldsymbol{A}^{\pm}=\boldsymbol{R}\boldsymbol{\Lambda}^{\pm}\boldsymbol{R}^{-1}$, 回到原始变量$\boldsymbol{u}$, 我们有

$$(\partial_t \boldsymbol{u})^{*}=-(\boldsymbol{A}^{+}\boldsymbol{u}_L'+\boldsymbol{A}^{-}\boldsymbol{u}_R')+\left(\boldsymbol{R}\boldsymbol{I}^{+}\boldsymbol{R}^{-1}\boldsymbol{C}\boldsymbol{u}_L+\boldsymbol{R}\boldsymbol{I}^{-}\boldsymbol{R}^{-1}\boldsymbol{C}\boldsymbol{u}_R\right)+\boldsymbol{D}. \tag{69}$$

特别当$\boldsymbol{u}_L=\boldsymbol{u}_R=\boldsymbol{u}^{*}$时有

$$(\partial_t \boldsymbol{u})^{*}=-(\boldsymbol{A}^{+}\boldsymbol{u}_L'+\boldsymbol{A}^{-}\boldsymbol{u}_R')+\boldsymbol{C}\boldsymbol{u}^{*}+\boldsymbol{D}. \tag{70}$$

从中可以看出如何应用 Lax-Wendroff 解法器和间断追踪计算$(\partial_t \boldsymbol{u})^{*}$。

多维情形是类似的. 例如考虑二维线性方程组

$$\begin{aligned} &\partial_t \boldsymbol{u}+\boldsymbol{A}\partial_x \boldsymbol{u}+\boldsymbol{B}\partial_y \boldsymbol{u}=\boldsymbol{C}\boldsymbol{u}+\boldsymbol{D}, \quad t>\boldsymbol{0}, \\ &\boldsymbol{u}(x,y,\boldsymbol{0})=\begin{cases}\boldsymbol{u}_L(x,y), & x<\boldsymbol{0}, \\ \boldsymbol{u}_R(x,y), & x>\boldsymbol{0},\end{cases} \end{aligned} \tag{71}$$

这里$\boldsymbol{A},\boldsymbol{B},\boldsymbol{C},\boldsymbol{D}$是常矩阵，且$\boldsymbol{A},\boldsymbol{B}$可对角化. 切向变化量$\boldsymbol{B}\partial_y \boldsymbol{u}$可看作相对法向的一个扰动，将线性方程组写成如下形式

$$\begin{aligned} &\partial_t \boldsymbol{u}+\boldsymbol{A}\partial_x \boldsymbol{u}=-\boldsymbol{B}\partial_y \boldsymbol{u}+\boldsymbol{C}\boldsymbol{u}+\boldsymbol{D}, \quad t>0, \\ &\boldsymbol{u}(x,y,0)=\begin{cases}\boldsymbol{u}_L(x,y), & x<0, \\ \boldsymbol{u}_R(x,y), & x>0.\end{cases} \end{aligned} \tag{72}$$

与上面类似方法，我们可以得到

$$\begin{aligned}(\partial_t \boldsymbol{u})^{*}=&-[\boldsymbol{A}^{+}(\partial_x \boldsymbol{u})_L+\boldsymbol{A}^{-}(\partial_x \boldsymbol{u})_R]-[\boldsymbol{R}\boldsymbol{I}^{+}\boldsymbol{R}^{-1}\boldsymbol{B}(\partial_y \boldsymbol{u})_L+\boldsymbol{R}\boldsymbol{I}^{-}\boldsymbol{R}^{-1}\boldsymbol{B}(\partial_y \boldsymbol{u})_R] \\ &+(\boldsymbol{R}\boldsymbol{I}^{+}\boldsymbol{R}^{-1}\boldsymbol{C}\boldsymbol{u}_L+\boldsymbol{R}\boldsymbol{I}^{-}\boldsymbol{R}^{-1}\boldsymbol{C}\boldsymbol{u}_R)+\boldsymbol{D},\end{aligned} \tag{73}$$

这里记号与上述意思相同，并有$(\partial_x \boldsymbol{u})_L=\lim\limits_{x\to 0-}\partial_x \boldsymbol{u}(x,y)$，$(\partial_x \boldsymbol{u})_R=\lim\limits_{x\to 0+}\partial_x \boldsymbol{u}(x,y)$等。从(70)和(73)可以看出，所有的信息都是迎风且时空耦合的。

特别强调，通过 GRP 解法器把切向变化自然地蕴含到数值通量中，这是法向（一维）黎曼解法器做不到的【38】。换句话说，如果只使用黎曼问题解法器，导致格式失去多维性，该格式应用到多维问题模拟时自然会有数值缺陷，不得不用其它方式进行补偿。

### 4.2. 声波近似 （Acoustic Approximation）---线性化 GRP 解法器

对于非线性系统来说，当流场是光滑的或者波的强度不大时，使用近似解法器进行通量的逼近就可以取得不错的效果，即声波近似或线性化 GRP 解法器。Toro 等的 ADER 解法器就是 GRP 的声波近似版本【35】。

先看下面的一维双曲平衡律(hyperbolic balance laws)

$$\begin{aligned} &\boldsymbol{\partial_t u} + \partial_x \boldsymbol{F}(\boldsymbol{u}) = \boldsymbol{H}(x,\boldsymbol{u}), \quad x \in \mathbb{R}, \quad t > 0, \\ &\boldsymbol{u}(x,0) = \begin{cases} \boldsymbol{u}_L(x), & x < 0, \\ \boldsymbol{u}_R(x), & x > 0. \end{cases} \end{aligned} \tag{74}$$

可以把它看作描写相对于网格界面法向效应的方程。如果$\boldsymbol{u}_L(0-0) = \boldsymbol{u}_R(0+0)$， 而$\boldsymbol{u}_L'(0-0) \neq \boldsymbol{u}_R'(0+0)$， 只有线性波(声波)从(0,0)点发出. 记$\boldsymbol{A}(\boldsymbol{u}) = \partial_{\boldsymbol{u}} \boldsymbol{F}(\boldsymbol{u})$，$\boldsymbol{u}^* = \boldsymbol{u}_L(0-0) = \boldsymbol{u}_R(0+0)$，$\boldsymbol{u} = \boldsymbol{u}^* + \boldsymbol{v}$，$\boldsymbol{v}$可看成$\boldsymbol{u}$在$\boldsymbol{u}^*$附近的扰动。然后在$\boldsymbol{u}^*$处线性化(73)可得

$$\boldsymbol{\partial_t v} + \boldsymbol{A}(\boldsymbol{u}^*)\boldsymbol{\partial_x v} = \boldsymbol{H}(x,\boldsymbol{u}^*), \quad \boldsymbol{u} = \boldsymbol{u}^* + \boldsymbol{v}, \tag{75}$$

这是一个线性系统。类似于(66)，从中可以计算$(\partial_t \boldsymbol{u})^* = (\partial_t \boldsymbol{v})^*$,

$$(\boldsymbol{\partial_t v})^* = -\boldsymbol{A}^+(\boldsymbol{u}^*)\boldsymbol{v}_L'(0) - \boldsymbol{A}^-(\boldsymbol{u}^*)\boldsymbol{v}_R'(0) + \boldsymbol{H}(0,\boldsymbol{u}^*), \tag{76}$$

这里$\boldsymbol{A}^\pm(\boldsymbol{u}^*) = \boldsymbol{R}(\boldsymbol{u}^*)\boldsymbol{\Lambda}^\pm(\boldsymbol{u}^*)\boldsymbol{R}^{-1}(\boldsymbol{u}^*)$，$\boldsymbol{\Lambda}^\pm(\boldsymbol{u}^*)$是由$\boldsymbol{\Lambda}(\boldsymbol{u}^*)$的“±”部分组成$\boldsymbol{\Lambda}^+ = diag\{\max(\lambda_i, 0)\}$，$\boldsymbol{\Lambda}^- = diag\{\min(\lambda_i, 0)\}$。从而

$$(\partial_t \boldsymbol{u})^* = -\boldsymbol{A}^+(\boldsymbol{u}^*)\boldsymbol{u}_L'(0) - \boldsymbol{A}^-(\boldsymbol{u}^*)\boldsymbol{u}_R'(0) + \boldsymbol{H}(x,\boldsymbol{u}^*). \tag{77}$$

这就是线性化 GRP 解法器[8].

当$\boldsymbol{u}_L(0-0) \neq \boldsymbol{u}_R(0+0)$时，声波近似策略如下：以局部黎曼解$\boldsymbol{u}^* = R(0; \boldsymbol{u}_L(0), \boldsymbol{u}_R(0))$为背景解，线性化(74)得到(75)，从而可得线性化 GRP 解法器(77),具体见【32】。

对于多维系统（仍然以二维为例），

$$\begin{aligned} &\boldsymbol{\partial_t u} + \partial_x \boldsymbol{F}(\boldsymbol{u}) + \partial_y \boldsymbol{G}(\boldsymbol{u}) = \boldsymbol{H}(x,y,\boldsymbol{u}), \quad t > 0, \\ &\boldsymbol{u}(x,y,0) = \begin{cases} \boldsymbol{u}_L(x,y), & x < 0, \\ \boldsymbol{u}_R(x,y), & x > 0, \end{cases} \end{aligned} \tag{78}$$

采用声波近似的策略，线性化这个方程组得到

$$\begin{aligned} &\boldsymbol{\partial_t v} + \boldsymbol{A}(\boldsymbol{u}^*)\partial_x \boldsymbol{v} = -\boldsymbol{B}(\boldsymbol{u}^*)\partial_y \boldsymbol{v} + \boldsymbol{H}(x,y,\boldsymbol{u}^*), \quad t > 0, \\ &\boldsymbol{v}(x,y,0) = \begin{cases} \boldsymbol{v}_L(x,y), & x < 0, \\ \boldsymbol{v}_R(x,y), & x > 0, \end{cases} \end{aligned} \tag{79}$$

---

[8] 当源项$\boldsymbol{H}(x,\boldsymbol{u})$存在时，即使$\boldsymbol{u}_L' = \boldsymbol{u}_R' \equiv \boldsymbol{0}$，$(\partial_t u)^* \neq 0$，意味着 GRP 解法器仍然起着重要作用。

从而得到类于(73)的解法器，这里$\boldsymbol{B}(\boldsymbol{u}^*) = \partial_{\boldsymbol{u}}\boldsymbol{G}(\boldsymbol{u})$。

### 4.3. 真正非线性 GRP 解法器

在(74)中，如果初始数据在原点(相对网格尺寸)有“很大”跳跃，就会出现强非线性波，线性化 GRP 解法器不足以分辨强非线性波，这时候需要使用真正非线性 GRP 解法器，否则就会出现明显的数值缺陷【7,34,35】. 一般的界定标准为

$$\| \boldsymbol{u}_L(\boldsymbol{0}) - \boldsymbol{u}_R(\boldsymbol{0}) \| \gg \boldsymbol{\alpha}\boldsymbol{\Delta}\boldsymbol{x}, \tag{80}$$

参数$\alpha > 0$是一个重要的经验参数。求解 GRP 解法器的基本思想是：解析强稀疏波，追踪强间断。特别需要指出，在强间断的情形下，热力学效应起着重要作用，真正非线性的 GRP 解法器反映了热力学相容的性质【7】。

本文不具体给出真正非线性 GRP 解法器.对于可压缩欧拉方程组，可参阅【5，36】，后期发展的不依赖于坐标系统的 GRP 解法器，详见【6】；对一般的双曲方程组，可参见【37】。

对于多维情形，仍然可以把切向的变化和间断面的变化看作扰动，考虑有下列形式问题

$$\begin{aligned} &\boldsymbol{\partial_t u} + \partial_x \boldsymbol{F}(\boldsymbol{u}) = -\partial_y \boldsymbol{G}(\boldsymbol{u}) + \boldsymbol{H}(x, y, \boldsymbol{u}), \qquad t > 0, \\ &\boldsymbol{u}(x, y, 0) = \begin{cases} \boldsymbol{u}_L(x, y), & x < 0, \\ \boldsymbol{u}_R(x, y), & x > 0. \end{cases} \end{aligned} \tag{81}$$

从而利用4.1和4.2节中法向解法器求解(72)和(73)，这主要源自于一个关键的事实：切向扰动在法向的传播是连续的。由此切向的变化在通量中得到充分反映，得到的数值方法具有真正的多维性【38】。

正如前面所讨论的那样，除非涉及（依赖于网格移动）中心拉格朗日型数值方法，这里不关注顶点多维黎曼解法器。由于真正多维黎曼问题的理论基础很不成熟【27,39,40,41】，真正多维顶点黎曼解法器常常带来不稳定的数值结果【42】。在实践中，人们更倾向使用健壮的逼近 GRP 解法器，例如，Maire 等利用新的框架并结合拉格朗日型 GRP 解法器【43】，发展了稳定的中心型高阶拉氏方法。

### 4.4. 一般时空关联模型的 GRP 解法器

对于一般的时空关联模型，如多相流、多介质模型【44】和 Navier-Stokes 方程等，广义黎曼问题解法器可以进行类似的构造，现进行简述.

对于多相流、多介质模型，广义黎曼问题解法器可以归结为一般的非线性平衡律的范畴，模拟的时空关联性质在 GRP 解法器中可以得到充分体现【45,46,47】。困难在于由介质组份以及混合引起的数值振荡、物质分数为负等非物理解，涉及物理建模本身的缺陷以及有限体积格式的投影(平均)过程。由于热力学关系的高度非线性，投

影过程不能反映精确的物理过程，例如内能的平均过程导致物质界面附近的压力振荡.因此，在投影步实现时空耦合，充分利用解的信息也许可以提高相关数值质量。

对于 Navier-Stokes 方程组等宏观模型，基本的想法是类似的。对于对流项（欧拉部分），使用上述标准的广义黎曼解法器。需要指出的是：初始数据(41)需要进行适当的匹配。也就是说该初始数据至少是二阶以上分片多项式。另外可使用的策略为：对于对流占优问题，采用松弛策略，把相应模型双曲化【48，49】。这样的策略是基于能量原理：在对流占优情形下，能量集中于初始能量传播影响区域内（双曲性质）。

从 Boltzmann 方程直接出发的动理学方法(Kinetic methods)，是设计流体力学数值方法的一条有效途径. 一般来说，很难写出 Boltzmann 方程的解，但在特定的近似假定下，如 BGK 模型【50】，可以类似于(58)那样隐式得出解的表达式，并将之用于数值通量的构造，得出的数值方法如气体动理学格式(GKS)【9】、统一气体动理学格式 UGKS【10】等。特别地，GKS 可以用来模拟 Navier-Stokes(N-S)方程描述的流动，可以作为 Navier-Stokes 的解法器来使用。按照这样的定义，在通量的构造部分，GKS 和 UGKS 解法器显然是时空耦合的。

### 4.5. 高精度方法中黎曼解法器和 GRP 解法器的比较

在高精度数值方法中，黎曼解法器和 GRP 解法器都可用来构造数值通量。前者在高阶线方法(high order methods of line)中常用，如 Runge-Kutta 型高阶方法，后者用在两步四阶方法中。下面仍以一维双曲守恒律方程组(31)来比较两种解法器的差别。考察控制体$\left[x_{j-\frac{1}{2}},x_{j+\frac{1}{2}}\right]\times[t_n,t_{n+1})$边界上通量。在$x=x_{j+\frac{1}{2}}$，记

$$\boldsymbol{u}_{j+\frac{1}{2}}^n=\lim_{t\to t_n+}\boldsymbol{u}\left(x_{j+\frac{1}{2}},t\right),\qquad (\partial_t\boldsymbol{u})_{j+\frac{1}{2}}^n=\lim_{t\to t_n+}\partial_t\boldsymbol{u}\left(x_{j+\frac{1}{2}},t\right). \tag{82}$$

由解关于时间$x$的正则性得到

$$\boldsymbol{F}\left(\boldsymbol{u}\left(x_{j+\frac{1}{2}},t\right)\right)=\boldsymbol{F}\left(\boldsymbol{u}_{j+\frac{1}{2}}^n\right)+\boldsymbol{F}'\left(\boldsymbol{u}_{j+\frac{1}{2}}^n\right)(\partial_t\boldsymbol{u})_{j+\frac{1}{2}}^n(t-t_n)+\mathcal{O}((t-t_n)^2). \tag{83}$$

（i） 黎曼解法器. 选取$\boldsymbol{F}_{j+\frac{1}{2}}^{Godu}=\boldsymbol{F}\left(\boldsymbol{u}_{j+\frac{1}{2}}^n\right)$，从而有

$$\begin{aligned}&\int_{t_n}^{t_{n+1}}\left[\boldsymbol{F}\left(\boldsymbol{u}\left(x_{j+\frac{1}{2}},t\right)\right)-\boldsymbol{F}\left(\boldsymbol{u}\left(x_{j-\frac{1}{2}},t\right)\right)\right]dt-\Delta t_n[\boldsymbol{F}_{j+\frac{1}{2}}^{Godu}-F_{j-\frac{1}{2}}^{Godu}]\\&=\frac{(\Delta t_n)^2}{\boldsymbol{2}}\left(\boldsymbol{F}'\left(\boldsymbol{u}_{j+\frac{1}{2}}^n\right)(\partial_t\boldsymbol{u})_{j+\frac{1}{2}}^n-\boldsymbol{F}'\left(\boldsymbol{u}_{j-\frac{1}{2}}^n\right)(\partial_t\boldsymbol{u})_{j-\frac{1}{2}}^n\right)+\mathcal{O}((\Delta t_n)^3)\end{aligned} \tag{84}$$

所以对于间断解来说，通量的截断误差为

$$E_j^n = \mathcal{O}((\Delta t_n)^2) \tag{85}$$

误差阶$q = 0$。不过，对于解的光滑区域来说，(84)中的差赋予了一个“阶数”，

$$E_j^n = \mathcal{O}((\Delta t_n)^3) \tag{86}$$

从而误差阶为$q = 1$。

（ii） GRP 解法器．选取$\boldsymbol{F}_{j+\frac{1}{2}}^{GRP} = \boldsymbol{F}\left(\boldsymbol{u}_{j+\frac{1}{2}}^n\right) + \frac{(\Delta t_n)^2}{\mathbf{2}}\boldsymbol{F}'\left(\boldsymbol{u}_{j+\frac{1}{2}}^n\right)(\partial_t \boldsymbol{u})_{j+\frac{1}{2}}^n$，从而有

$$\int_{t_n}^{t_{n+1}} \left[\boldsymbol{F}\left(\boldsymbol{u}\left(x_{j+\frac{1}{2}}, t\right)\right) - \boldsymbol{F}\left(\boldsymbol{u}\left(x_{j-\frac{1}{2}}, t\right)\right)\right] dt - \Delta t_n [\boldsymbol{F}_{j+\frac{1}{2}}^{GRP} - F_{j-\frac{1}{2}}^{GRP}] \\ = \frac{(\Delta t_n)^3}{6}\left(\partial_t^2 \boldsymbol{F}\left(\boldsymbol{u}_{j+\frac{1}{2}}^n\right) - \partial_t^2 \boldsymbol{F}\left(\boldsymbol{u}_{j-\frac{1}{2}}^n\right)\right) + \mathcal{O}((\Delta t_n)^4). \tag{87}$$

与上面讨论类似，对于间断解，GRP 解法器有一阶精度$q = 1$，而对于光滑解有二阶精度$q = 2$。

### 4.6 时空耦合数值边界条件

边界条件的数值处理是计算流体力学中的一个核心问题，大部分的数值处理基于对虚拟区域的延拓或特征传播理论【51】。最近逆 Lax-Wendroff 方法【52】用来数值处理边界条件，这种方法蕴含了时空耦合性。我们认识到，流体力学方程组的数值边界条件等价于单边广义黎曼问题问题的数值求解(one-sided GRP solver)【53】。事实上，先前的工作已经认识到单边黎曼问题在数值边界条件处理上的重要性【54，55】。这些工作与直接的逆 Lax-Wendroff 方法相比，有以下特点：

*(i)* 在有限体积框架下，计算区域的边界自然为控制体的边界，所以数值边界条件处理等价于边界上的通量数值逼近，它归结为单边黎曼解法器的构造。

*(ii)* 单边黎曼解法器中的$(\partial_t u)^*$其实反映了边界条件上的受力情况，即牛顿第二定律的数学表现，这是时空耦合算法的体现。在【53】中，这一事实是单边黎曼解法器的基石。

*(iii)* 在技术上，如此处理数值边界条件可以尽量少使用虚拟网格。在时空二阶格式中无需使用虚拟网格； 相比较其它更高阶方法，至多使用一半的虚拟网格。

*(iv)* 无需使用更高阶导数，避免了人为的复杂性和虚假物理性质等【52】。

单边黎曼解法器的使用将极大简化了边界条件的数值处理，特别是解决了无结构网格下虚拟区域的插值问题【56】。

### 4.7 高阶时空耦合算法

高阶精度数值方法对小尺度问题的数值模拟非常重要，如湍流等。基于有限体积格式的框架(25)，高阶方法需要在数值通量和数据重映两部分具有高阶逼近性质．在通

量的逼近部分，可以采用Lax-Wendroff方法，但流体力学方程组(8)的非线性性质导致高阶展开过于复杂，缺乏实际工程价值。更糟糕的是，解的间断性质意味着高阶展开的运算没有意义， 因此需要寻求其它方式。在过去的几十年中，各种空间重构技术以及 Runge-Kutta 型时间推进方法取得了巨大的进步，读者可以容易查阅，这里不再赘述。下面简述一类时空耦合方法。

文【29】发展了两步四阶时间推进方法，成功地融合了 Runge-Kutta 型方法简单性优点和基于 GRP 解法器的时空耦合性质。文【4】详述了该类方法的特性： (i)计算效率：同等条件下其计算效率是 Runge-Kutta 方法的一半（二维情形为例）【57】; (ii)紧致性：由于时间推进步骤减少一半，模版就会减少一半；时空耦合的 HWENO 型重构【22, 58】 可以使计算格式更加紧致; (iii) 稳定性：已经证明其稳定区域比 Runge-Kutta 大【59】; (iv) 兼容性：该方法很容易和其它方法相结合，如 GKS 解法器【60】、多矩方法【61】以及 CRP 重构技术【62】等。

目前，该类方法还局限在“显式”框架内，相关的研究处于起步阶段。为了应用的需要，亟须发展“隐式”高阶方法，以适应“强刚性”等多尺度问题的求解。

## 五 数值算例

在可压缩流体的算法设计及其数值模拟中，有大量基准算例(Benchmarks)。随着算法进步和工程需求的提高,基准算例应该与时俱进，面对相当极端的环境. 在【63】中，一系列基准算例值得新的数值算法测试。下面几个算例，可以看出时空耦合性的重要性。

### 算例 5.1. 大压力比（大密度比）问题

这个问题又称为 LeBlanc 问题，它是经典的激波管问题的极端情形，从中可以看出数值方法对强稀疏波的分辨以及对激波的影响。控制方程为一维欧拉方程（31），初始数据具有如下形式

$$(\rho,u,p)(x,0)=\begin{cases}(10^4,0,10^4), & 0\le x<0.3,\\ (1,0,1), & 3\le x\le 1.0.\end{cases} \tag{88}$$

多方指数取$1.4$，计算的终止时间$t=0.12$。文【64】利用这个例子检测了多个数值格式的性能。这里我们同样用这个算例针对性讨论本文涉及的一些观点，数值结果来源于【7,29】。

首先在图 5.1(A)中展现使用不同解法器的一阶格式。可以看出一阶 Godunov 方法随着网格加密，可以收敛到精确解；使用逼近解法器，收敛相对较慢，但仍然具有收敛性. 图 5.1(B)使用了空间二阶重构，而时间离散使用一阶向前欧拉方法，解法器分别使用黎曼解法器和逼近的黎曼解法器(HLLC, Roe)， 数值表现很差，不具有收敛性。正如在 4.5 节所述，当空间具有高阶精度，使用一阶黎曼解法器构造数值通量等，算法不具有相容性;类似的情形反映在图 5.1(C)， 即使时间离散使用二阶 Runge-Kutta 方法。

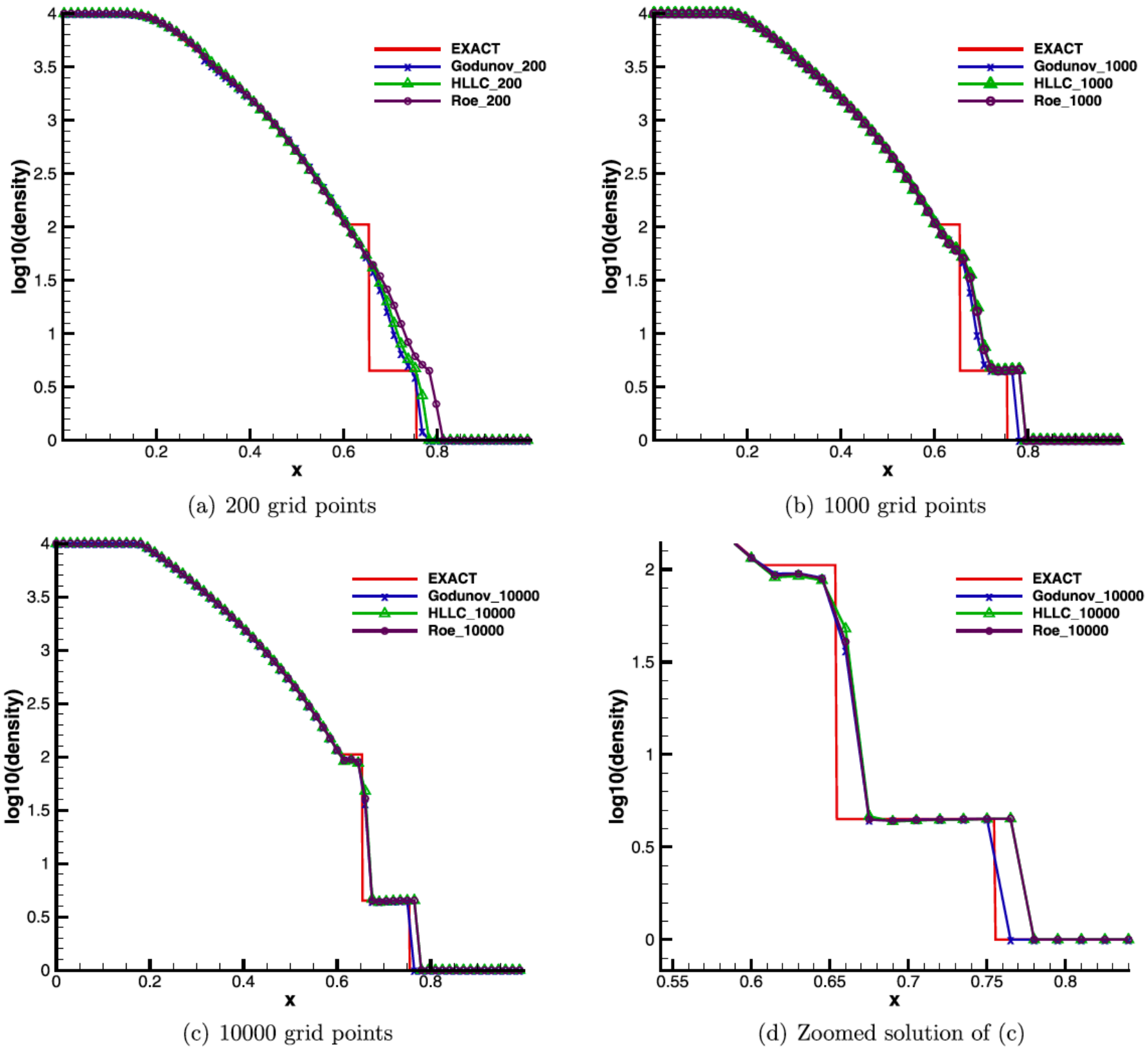

(a) 200 grid points (b) 1000 grid points

(c) 10000 grid points (d) Zoomed solution of (c)

图 5.1(A) 一阶格式的数值结果. 图标 Godunov, HLLC, Roe 分别表示不同的解法器

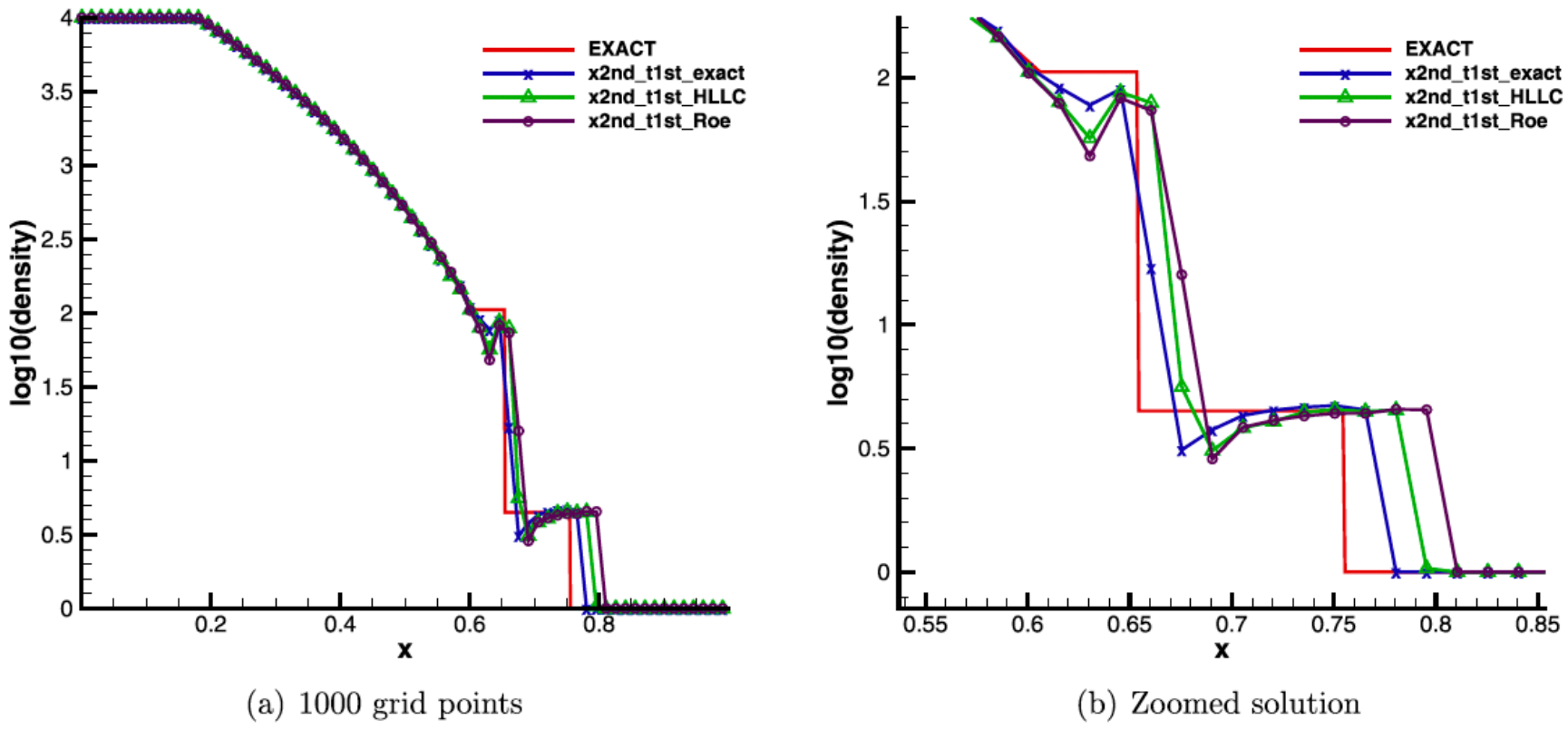

(a) 1000 grid points (b) Zoomed solution

图 5.1(B) 空间二阶时间一阶的数值结果. 间断处格式不具有收敛性

图 5.1(D)考察了在这种极端情况下使用线性化方法的数值表现，黎曼解使用声波近似解法器。其实【34,35】已经指出线性化解法器的数值缺陷。图 5.1(E)使用了非线性 GRP 解法器，数值结果立即改善，说明了强非线性出现后真正非线性 GRP 解法器的重要性。

图 5.1(F)展示的结果是基于 GRP 解法器的两步四阶方法【29】，从中看到用 100 网格点可以很好分辨间断，300 网格点可以得到完美效果。这是许多数值方法很难达到的，从而说明了时空耦合算法的必要性。

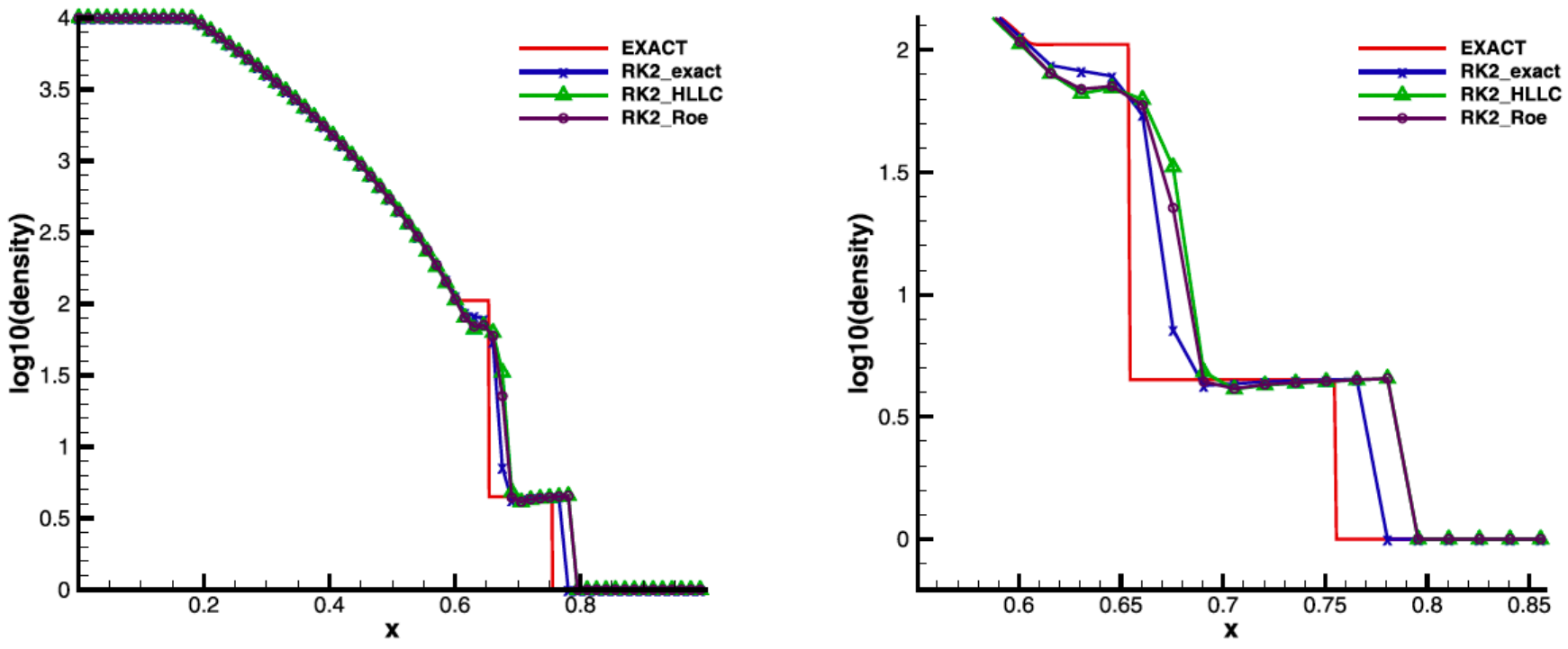

(a) 1000 grid points (b) Zoomed solution

图 5.1(C) Runge-Kutta 型二阶方法，使用精确和逼近黎曼解法器构造数值通量

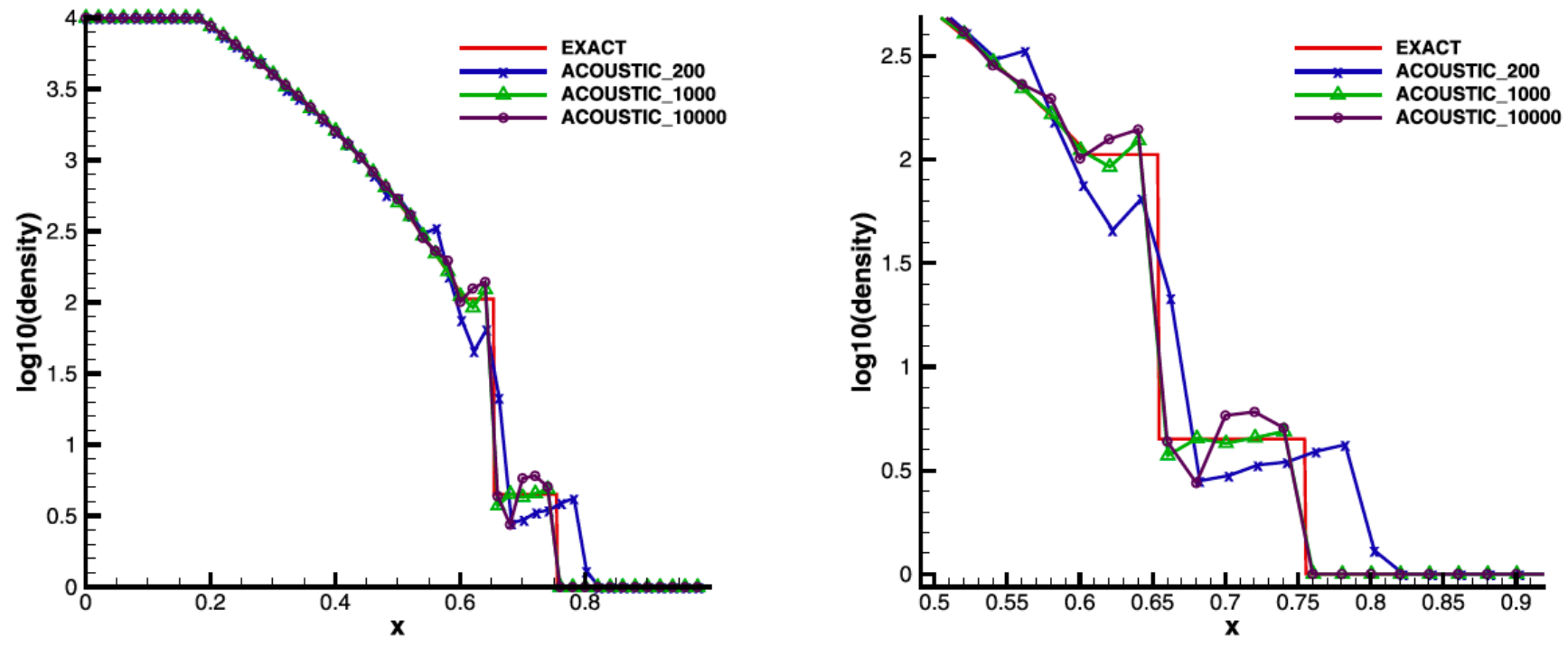

(a) Acoustic GRP solver with different grid points (b) Zoomed solution

图 5.1(D) 单步声波近似的 GRP 方法

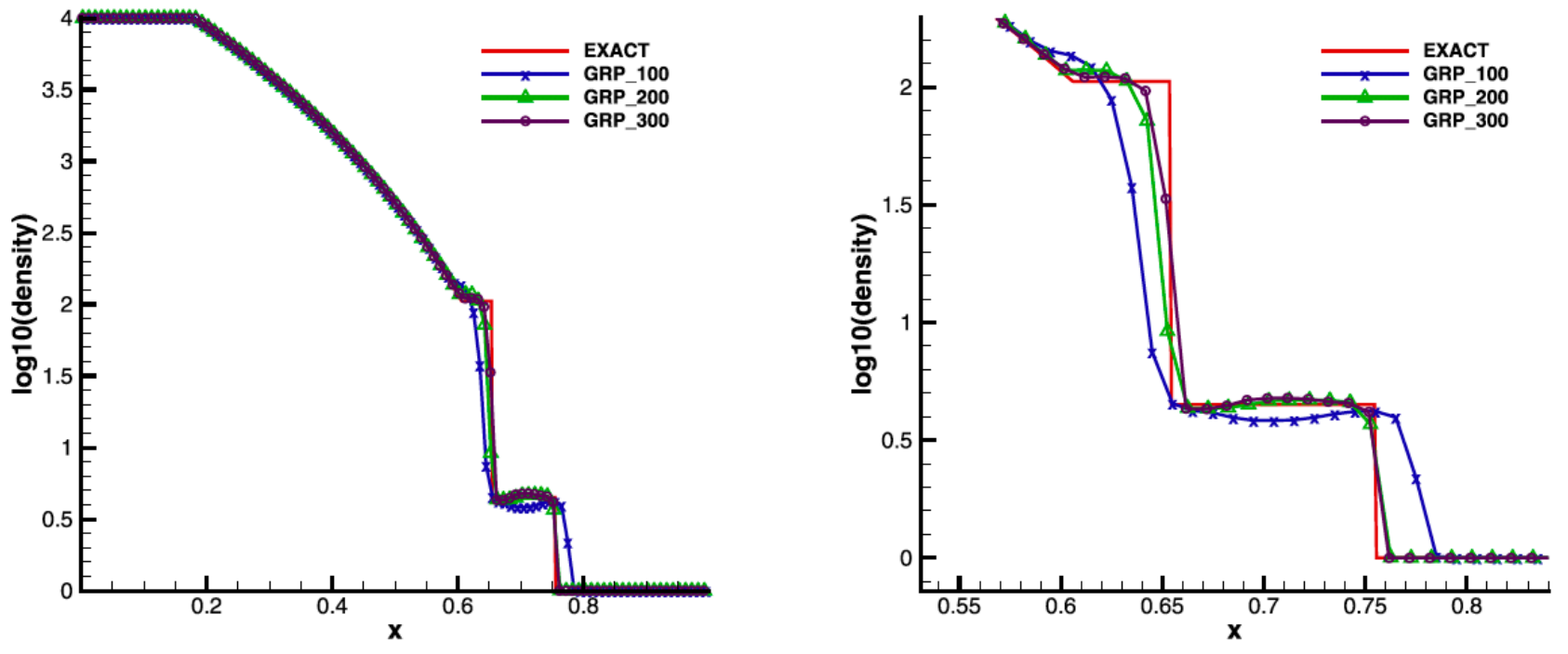

(a) GRP with relatively small number of grid points (b) Zoomed solution

图 5.1(E) 二阶 GRP 格式，使用了真正非线性 GRP 解法器

**算例 5.2. 激波和熵波相互作用的问题**

这个算例是 Shu-Osher 算例的推广，考虑了更极端的情形，又称为 Titarev-Toro 问题【62】。控制方程依然为欧拉方程，初始数据为

$$(\rho,u,p)(x,0)=\begin{cases}(1.515696,0.52336,1.805), & -5\le x<-4.5,\\ (1+\sin(20\pi x),0,1), & -4.5\le x\le 5.\end{cases} \tag{89}$$

这里使用了 GKS 解法器【9】得到图 5.2 的数值结果，详细说明见【60】。

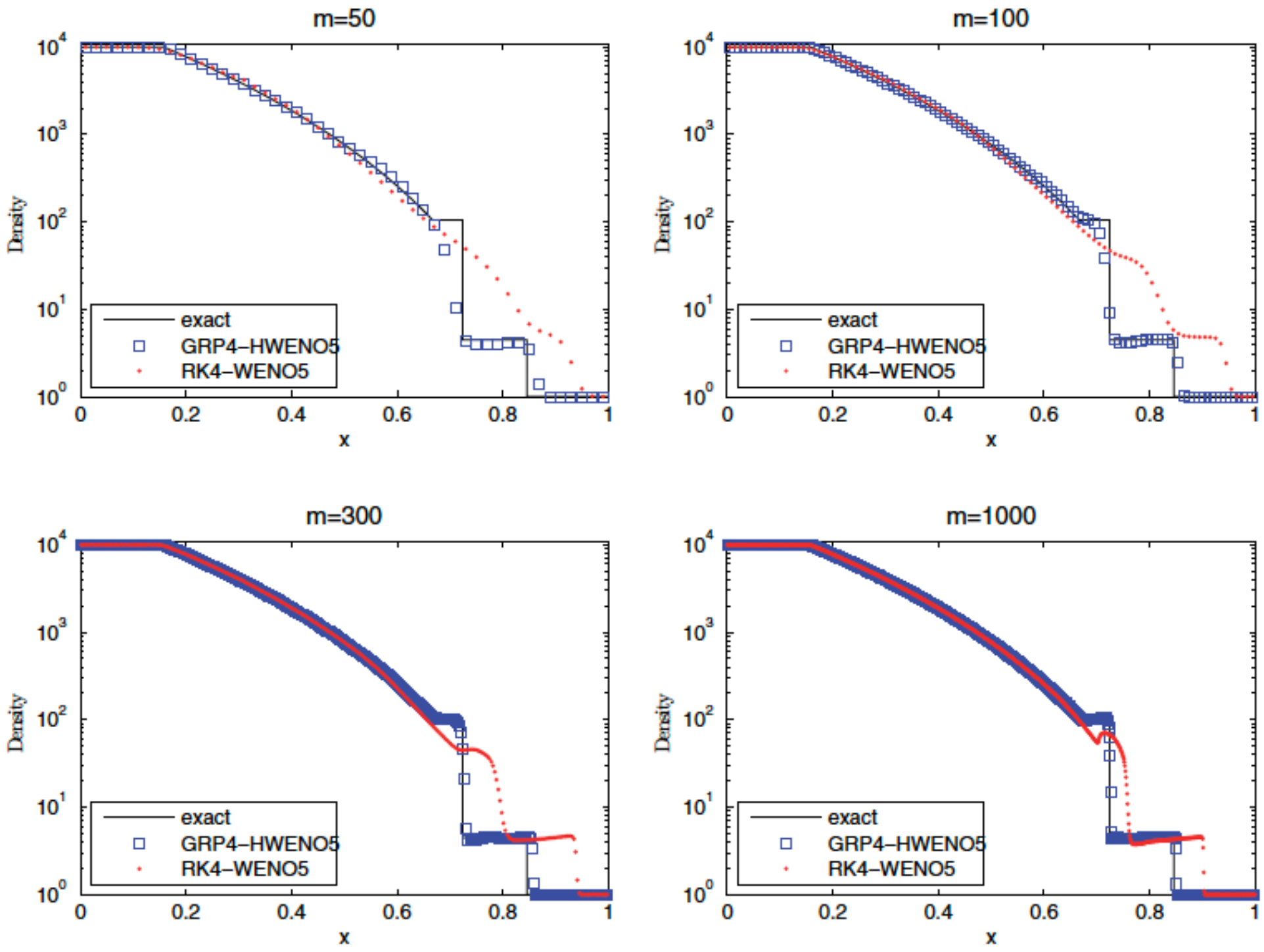

图 5.1(F) 高阶方法. 这里 m 表示所用的网格点数，RK4-WENO5 表示使用空间 5 阶 WENO 重构，4 阶 Runge-Kutta 时间推进方法；GRP4-HWENO5 表示使用空间 5 阶 HWENO 重构，两步四阶时间推进方法

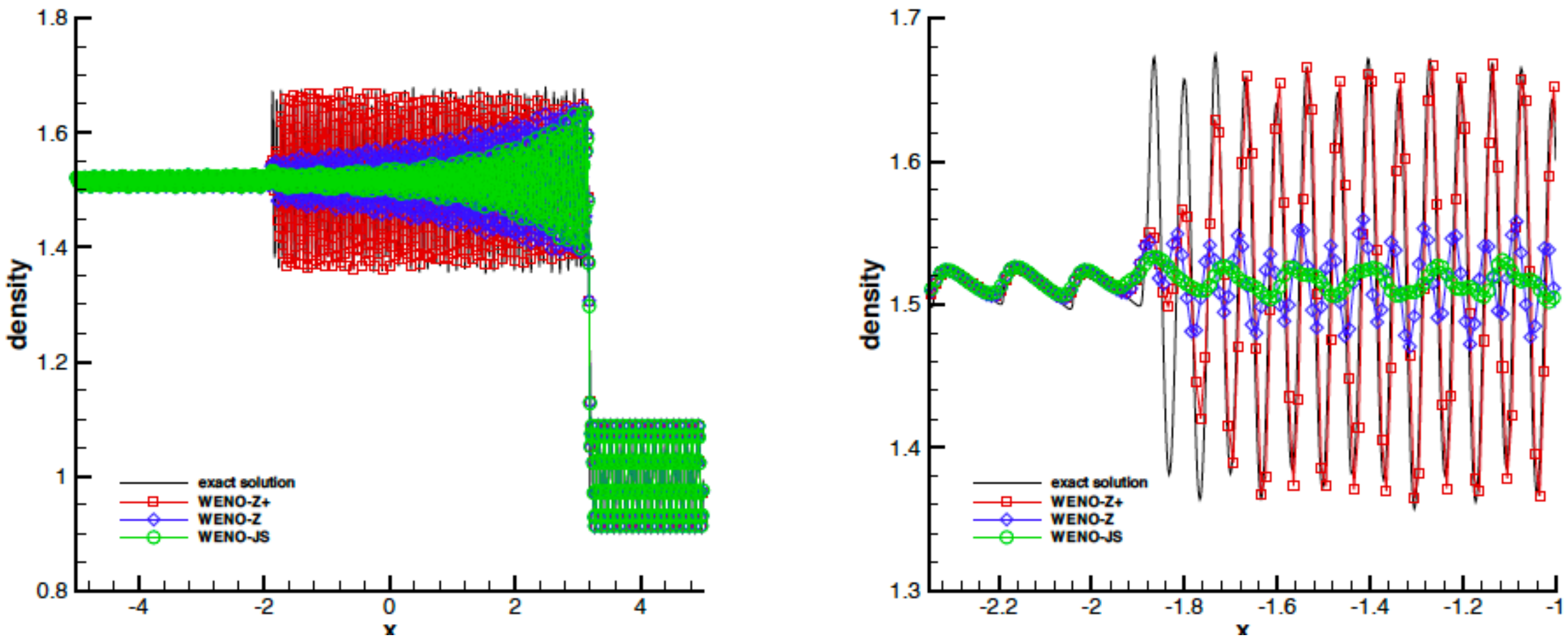

图 5.2. Titarev-Toro 问题.这里使用 1000 网格点数，空间重构采用了不同的重构策略，计算终止时间为$t = 5$

**算例 5.3. 激波和悬浮刚体的相互作用**

这个数值结果取自【54】，模拟了隧道$[0,1]\times[0,0.2]$内激波撞击一个长$a = 0.06$、宽$b = 0.03$矩形刚体的流场情况。控制方程为两维欧拉方程方程组，我们使用了时空耦合二阶 GRP 方法，并用单边广义黎曼解法器刻画方块移动时的数值边界条件。初始时刻一个马赫 3 的激波位于$x = 0.08$处，波前状态为$(\rho_0, u_0, v_0, p_0) = (1.3,0,0,0.1)$；该

方块的密度为$\rho_\Omega = 10\rho_0$，质量为$M_\Omega = ab\rho_\Omega$。初始时刻该方块以倾斜度$\frac{\pi}{4}$被放置于隧道中，重心为(0.15，0.06)，惯性力为$A_\Omega = \frac{M_\Omega}{3}(a^2+b^2)$。图 5.3 展示了两个时刻的流场压力分布情况。

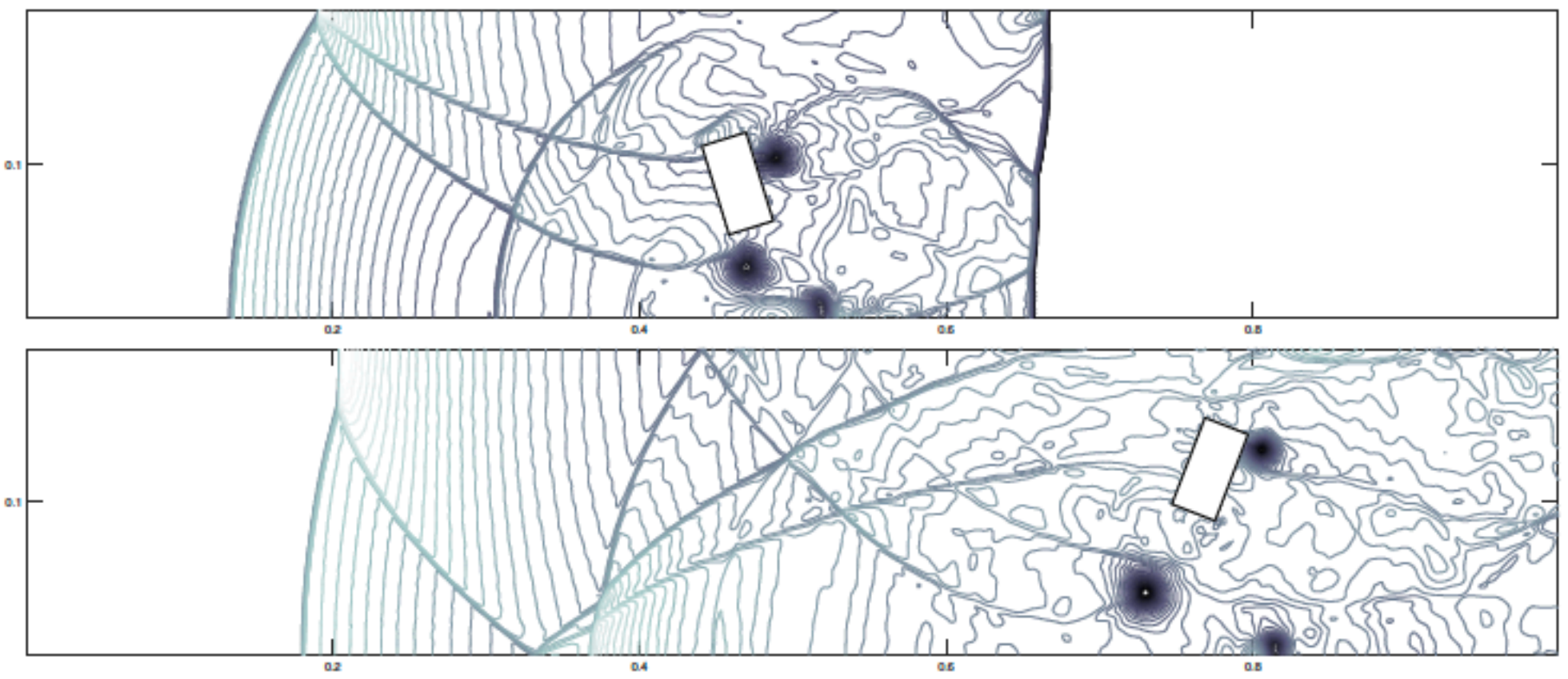

图 5.3. 激波和悬浮刚体的相互作用后压力的分布情况。这里使用$800\times160$个网格，上图计算终止时刻为$t=0.6$，下图为$t=1.0$

**算例 5.4. 多介质激波和起泡的相互作用**

这个例子展示了激波和气泡相互作用的情况，该问题已经成为多相流领域一个标准算例【66】。下面的结果取自【45】，分别用能量分裂的 Godunov 方法和 GRP 方法模拟所得结果。显然，GRP 很好提高了流场的分辨率。

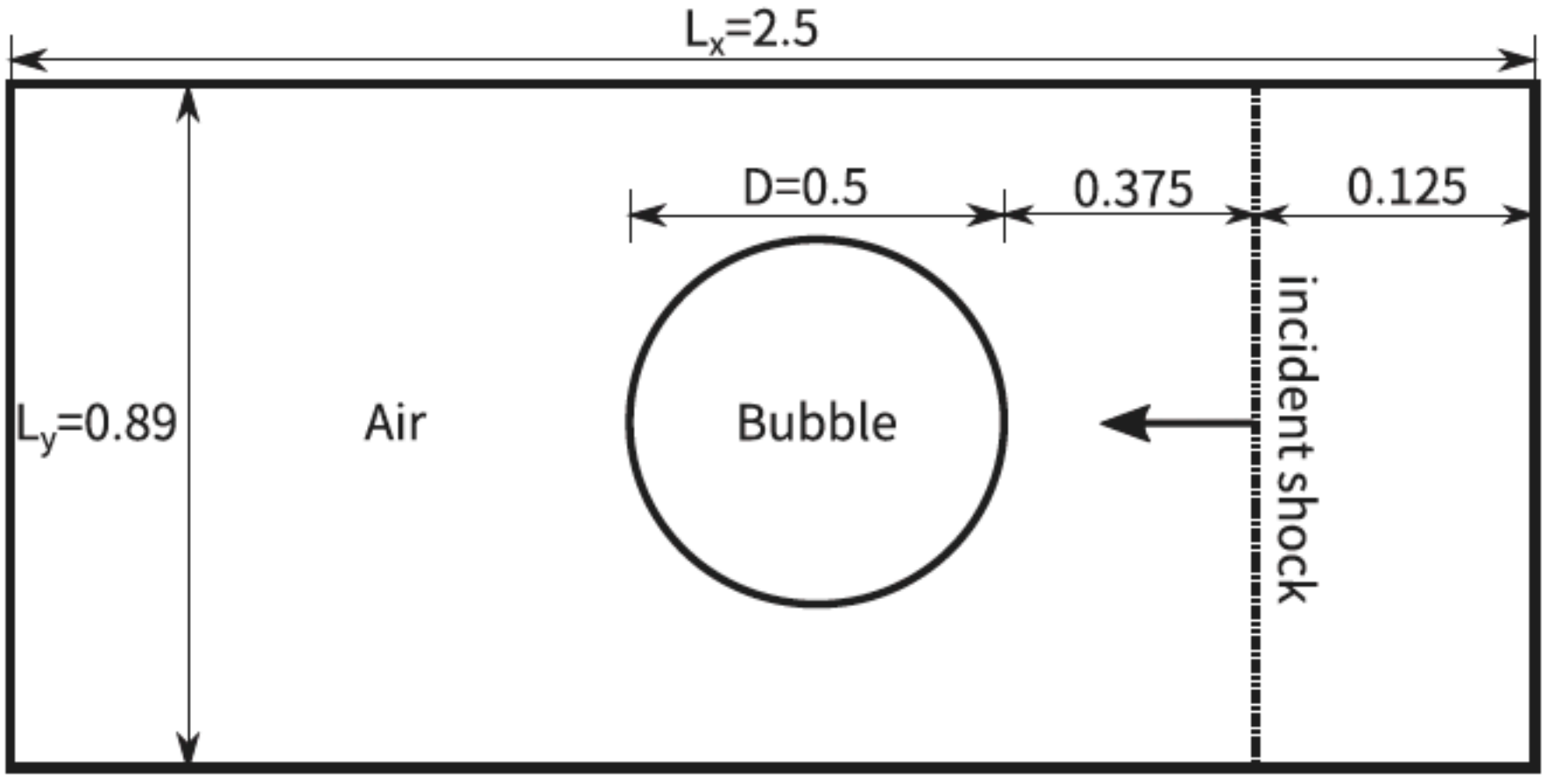

图 5.4(A). 激波撞击氦气泡的装置示意图。初始时刻激波的马赫数是$M_s = 1.22$，计算网格为$2500\times890$，其它参数详见【45】的算例 D

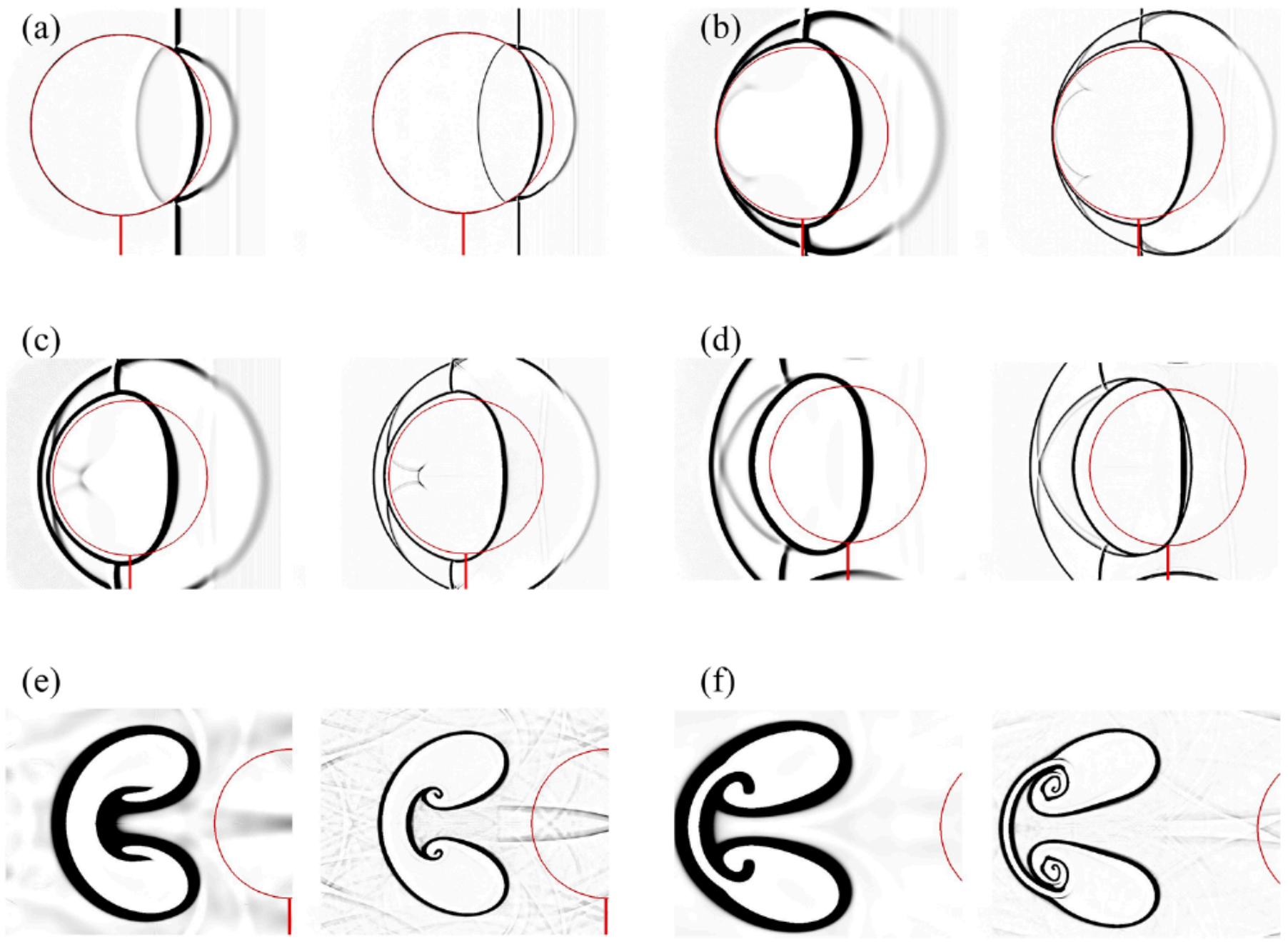

图 5.4(B).激波撞击气泡不同时刻的阴影图，每个子图的左边是 Godunov 格式的结果，右边是 GRP 的结果. (a) $t = 32\mu s$；(b) $t = 62\mu s$; (c) $t = 72\mu s$; (d) $t = 102\mu s$; (e) $t = 427\mu s$; (f) $t = 674\mu s$.

**算例 5.5. 各向同性可压缩湍流的模拟**

下面的算例采用了两步四阶 GKS 方法【60】模拟的超音速各向同性可压缩湍流的情况【67】，从中可查阅具体的计算参数，其中计算区域是$[-\pi,\pi]^3$。该文展示了两步四阶方法和二阶方法同样健壮。对许多高阶方法来说，分辨激波碎片(shocklet)是个巨大的挑战【68】，该方法中 GKS 解法器的时刻耦合性质继承了物理本身的属性。

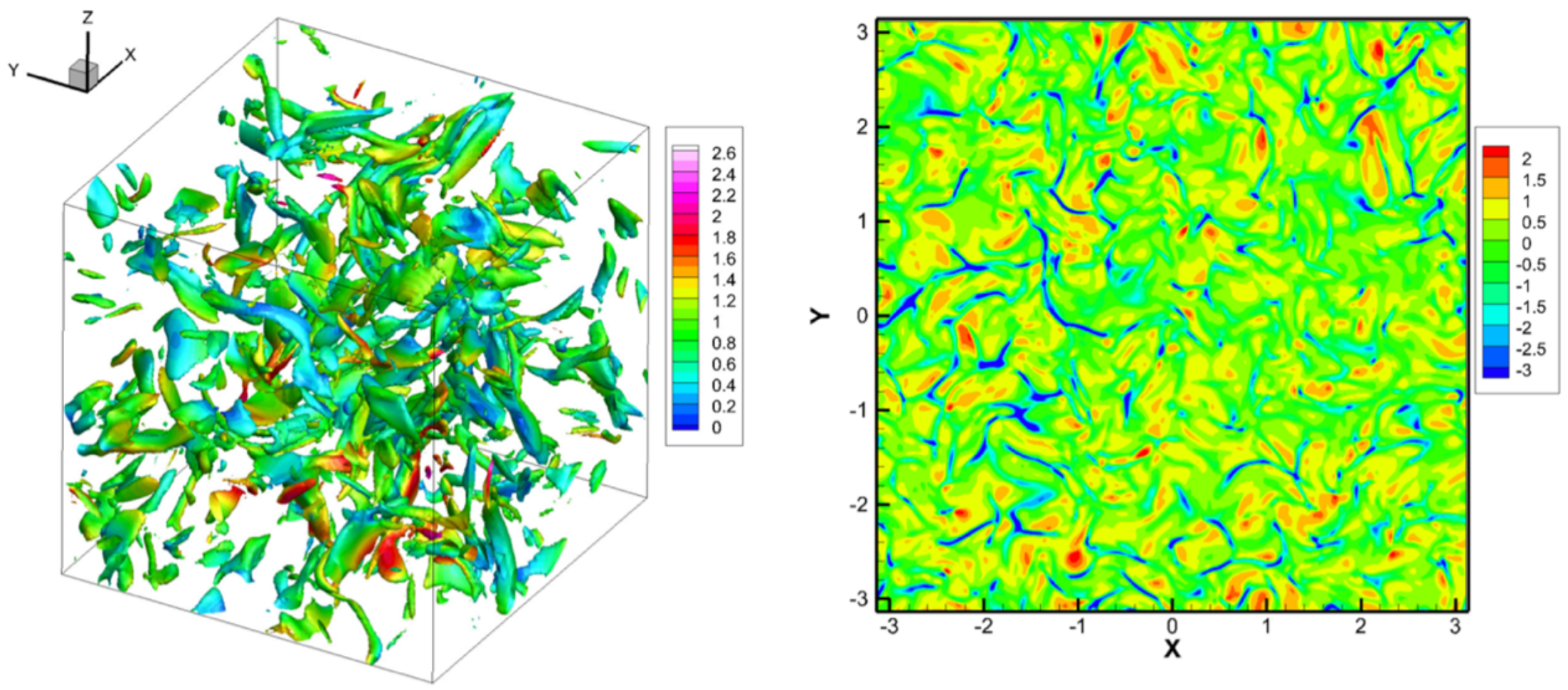

图 5.5. 各向同性可压缩湍流的模拟【68】。其中湍流马赫数$Ma_t = 1.2$, 初始泰勒微尺度雷诺数$Re_\lambda = 72$, 速度梯度张量第二不变量的等值面$Q$ =25

## 六 讨论与展望

流体力学的时空关联模型刻画了物理量空间变化在时间方向上的演化的传播，如各种波的传播等。当进行数值模拟时，这种性质理应得到继承，相应地所用算法应该具备时空耦合特性. 实践过程中这个看似简单课题理应得到关注。遗憾的是，相对于时空解耦方法，时空耦合算法需要更好理解模型的物理性质或数学理论，人们自然“避难就易”。一个直白的问题是：“即使物理问题的数学模型十分完美，当相应的算法缺乏相应完美性质时，其数值结果怎么令人信服呢？”

本文首先严格论述有限体积方法. 正如引言所解释的原因：(1)无论从物理定律的形成还是数值算法的构造，有限体积方法是解流体力学问题最自然的框架，本文总结了这个框架形成的数学理论和内在涵义；(2)流体现象的复杂性（如间断等）客观地阻碍了有限体积方法严格数学理论的形成，对于这一框架的认识常出现诸多似是而非的争论；(3)对于许多实际工程问题，基于无结构网格的算法设计是一个自然要求，在此之上许多方法相互冲突【24】，有必要从最基本的地方出发建立一个根本准则。幸运地是，许多工程软件，如 Fluent, 自动地在有限体积框架下生成；许多工程人员当遇到难以跨越的困难时，往往借用有限体积框架度过难关。从论证过程可以看到时空耦合是算法的根本属性。

有限体积格式的步骤很简单：数据重构（投影）和数值通量的构造。目前 CFD 算法的大部分研究者将注意力集中于前者，基本上在空间逼近论的范畴内进行探索。尽管黎曼不变量（Riemann invariants）等物理量以及其它技术用于重构过程，但是重要的时空耦合性质相当缺乏[9]。本文侧重于后者的讨论，给出了有限体积格式与积分平衡律(12)之间的相容性准则。特别对于数值通量介绍了黎曼解法器和 GRP 解法器，并在 4.5 节中进行了对比。简单地总结如下，具体内容可以到【2】中查阅。

（i） **关于 Riemann 解法器**. 对于双曲守恒律（欧拉方程），当初始数据是**分片常数**时。由 Riemann 解法器产生的数值通量是无穷阶相容，即 Godunov 格式就是积分平衡律；当初始数据是**非常数的分片光滑函数**时，Riemann 解法器产生的数值通量对光滑解是一阶相容的，但对间断解是不相容的，见 4.5 节。也就是说对于高阶空间重构，如果不能有匹配的数值通量，产生的数值格式是不相容的。值得注意的是，如果控制方程包含外力项时，Godunov 格式永远是不相容的。

（ii） **关于 GRP 解法器**. GRP 解法器给出时空耦合的数值通量.对于光滑流场，GRP 解法器得到二阶相容的数值通量；但对于间断解只有一阶时间精度。GRP 解法器是保证有限体积格式的逼近解收敛的一个必要条件。

这里之所以再次强调这点并细致给出算例 5.1，原因在于为了看清它与传统理解的差异。事实上，算法时空耦合性也体现在其它方面，比如最近研究气体动理学的渐近性质时，提出了统一保持性质（unified preserving property, UP）的概念【69】。对于一个动理学方法，不仅应该具有欧拉层面的渐近保持性质(Asymptotic preserving property,

---

[9] 在随机选取(Random Choice)方法中，黎曼问题问题的解被用在随机选取中. 在某种意义下，它可以看成是一种时空耦合投影方法。

AP)，还应该根据需要具有 Navier-Stokes 或更深层面的 UP 性质，这个过程实际上是通过时空耦合的思想实现的。该文分别用时空耦合 DUGKS 方法以及时空解耦 CLR 格式，对两维不可压缩 Taylor 涡的进行数值模拟，见图 6.1，具体见【69】。

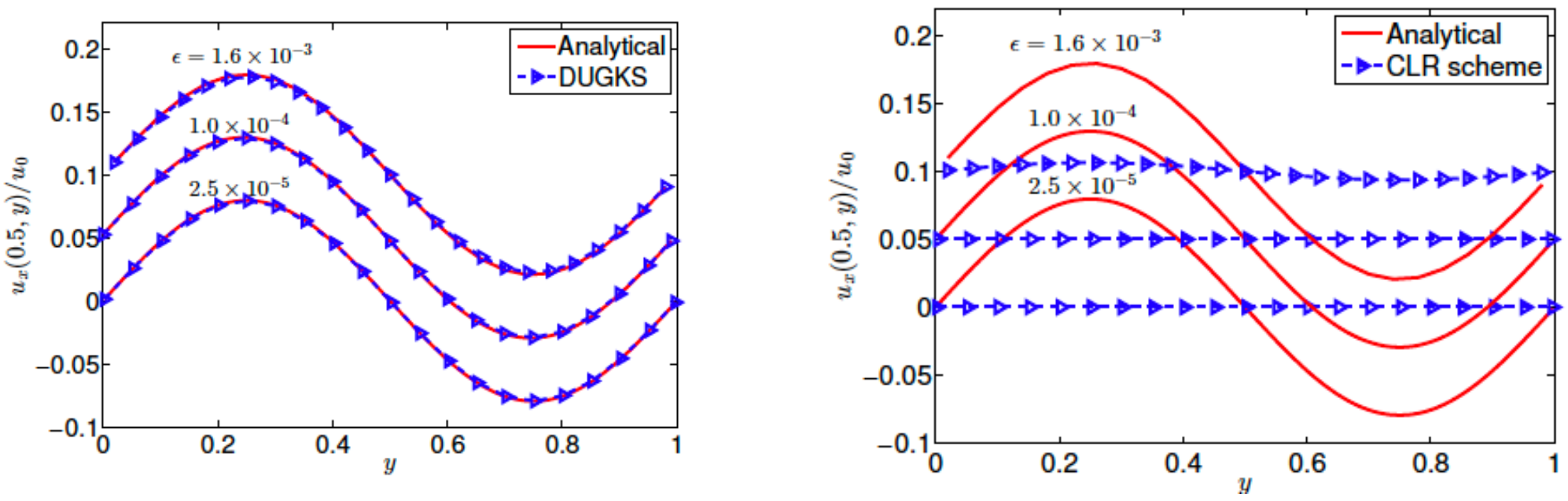

图 6.1. 关于二维不可压缩 Taylor 涡的 DUGKS(时空耦合)和 CLR(时空解耦)模拟的比较

由于篇幅限制，本文只给出了有限体积方法基本原理的相容性准则以及通过 GRP 解法器实现相容性的技术细节，对很多根本性质还没有涉及，如时空相容格式的熵稳定性等。对于可压缩流体力学来说，熵稳定性对应于热力学相容性【7】，这部分留在将来的文章中探讨。

客观地说，时空耦合算法的研究非常不成熟，目前只在相对比较“纯粹”的模型上进行分析和验证，但是这一思想应该加以推广与应用，比如如何将时空耦合算法的思想应用于一般的时空关联湍流模型【70】，粒子建模和时空耦合算法等。就算法本身来说，时空耦合的隐式格式研究还没有开展，这一领域的研究应该是下一阶段的重点。

在工程应用中时空耦合的程序显得更少，一方面是受限算法模块的历史延续性制约；另一方面是工程领域内的“理性”力学越来越少。加强工程算法的科学性是需要目前亟待解决的观念问题。

**后记.** 作为一个数学工作者，很忐忑地接受邀请在力学的专业刊物《空气动力学学报》奉献此类稿件，毕竟隔行如隔山；但想到数学、力学本就是“一家”，向力学家“学习”本就是“数学”的一大研究动机，心里稍微坦然点。

**致谢:** 此文是和我的下列合作者合作过程中完成的, 在此一并致谢: 他们是 Matania Ben-Artzi, 齐进、汪玥、杜知方、雷昕、徐昆、郭照立、李启兵等。特别感谢曹贵瑜博士提供的关于各向同性可压缩湍流模拟算例 5.5, 感谢徐昆教授、郭照立教授和汪玥副研究员提出许多根本性改进意见, 感谢陈海波给予了文字上的润色。

**A spacetime outlook on CFD: Spacetime correlated models and spacetime coupled algorithms**

Jiequan Li
Institute of Applied Physics and Computational Mathematics, Beijing;
Center for Applied Physics and Technology, Peking University

**Abstract:** A spacetime outlook on Computational Fluid Dynamics is advocated: models in fluid mechanics often have the spacetime correlation property, which should be inherited and preserved in the corresponding numerical algorithms. Starting from the fundamental formulation of fluid mechanics under continuum hypothesis, this paper defines the meaning of spacetime correlation of the models, establishes the fundamental principle of finite volume schemes, expounds the necessity of spacetime coupling of algorithms, as well as realizes the physical and mathematical unification of basic governing equations of fluid mechanics and finite volume schemes. In practice, the design methodology of spacetime coupling high order numerical algorithms is presented, and the difference from spacetime decoupling method is compared. It should be pointed out that most of the contents in this paper are suitable for computational fluid dynamics under the assumption of continuous medium, and some are only suitable for compressible flow.

**参考文献**